\documentclass[11pt,draftcls, onecolumn,journal]{IEEEtran}
\usepackage{times}
\usepackage[final]{graphicx}
\usepackage[reqno]{amsmath}
\usepackage{amsfonts}
\renewcommand{\baselinestretch}{0.94}
\usepackage{times,amsmath,epsfig}
\usepackage{latexsym,amssymb}
\usepackage{cite}
\usepackage{psfrag}
\usepackage{subfigure}
\usepackage{stfloats}
\usepackage{amsmath}
\usepackage{amssymb}
\usepackage{array}
\usepackage[noend]{algorithmic}
\usepackage{algorithm}
\usepackage{color}
\newtheorem{theory}{Theorem}

\def\myQED{\mbox{\rule[0pt]{1.5ex}{1.5ex}}}

\begin{document}
\baselineskip 18.5pt

\title{\LARGE Channel Exploration and Exploitation with Imperfect Spectrum Sensing in Cognitive Radio Networks}

\author{
Zhou Zhang, Hai Jiang\footnote{Correspondence author: Dr. Hai Jiang, Dept. of ECE, University of Alberta, Canada. Email: hai1@ualberta.ca.}, Peng Tan, and Jim Slevinsky
}
\maketitle
\pagestyle{plain}

\vspace{-8mm}\begin{abstract}
In this paper, the problem of opportunistic channel sensing and access in cognitive radio networks when the sensing is imperfect and a secondary user has limited traffic to send at a time is investigated. Primary users' statistical information is assumed to be unknown, and therefore, a secondary user needs to learn the information online during channel sensing and access process, which means learning loss, also referred to as regret, is inevitable. In this research, the case when all potential channels can be sensed simultaneously is investigated first. The channel access process is modeled as a multi-armed bandit problem with side observation. And channel access rules are derived and theoretically proved to have asymptotically finite regret. Then the case when the secondary user can sense only a limited number of channels at a time is investigated. The channel sensing and access process is modeled as a bi-level multi-armed bandit problem. It is shown that any adaptive rule has at least logarithmic regret. Then we derive channel sensing and access rules and theoretically prove they have logarithmic regret asymptotically and with finite time.  The effectiveness of the derived rules is validated by computer simulation.
\end{abstract}

\vspace{-5mm}\begin{flushleft} \textbf{\it Keywords} -- Cognitive radio; opportunistic channel access; bandit problem; channel exploration; channel exploitation.

\end{flushleft}
\thispagestyle{empty}

\section{Introduction}\label{s:intro}

Cognitive radio has emerged as an effective solution to alleviate the spectrum shortage problem and improve spectrum efficiency. It has received tremendous research attentions recently  \cite{JMitola, SHaykin, AGoldsmith, FMolisch, RZhang, HJiang}.
In a cognitive radio network, opportunistic spectrum access (OSA) is used, in which the unlicensed users, referred to as {\it secondary users}, search for {\it spectrum holes} through sensing, and utilize the observed spectrum opportunities for their data transmission. Optimal OSA when the secondary users have statistical information of licensed users (referred to as {\it primary users}), such as information of free probabilities of primary channels, has been addressed in \cite{CLiang, RFan,FanTVT2011, RZhang_1, QZhi}, to maximize transmission capacity, optimize transmission power efficiency, etc. However, research on the optimal OSA without {\it a priori} statistical knowledge of primary channels is still in its infancy. The research challenge is how to achieve the optimal tradeoff between channel exploration (the process to sense the channels so as to learn the statistical information) and channel exploitation (the process to utilize observed channel opportunities). If statistical information of primary channels is known in advance, a secondary user can select the optimal channels to sense and subsequently access sensed-free channels. However, without such information, a learning process is needed, and the secondary user should also explore suboptimal channels through sensing to learn statistical information of those channels. Therefore, learning loss is expected, compared to the case that the secondary user always selects the optimal channels. In the literature, the channel sensing and access process has been modeled as a multi-armed bandit problem (MABP) \cite{LLai}. For an MABP, the loss due to learning until time instant $t$ is represented by the {\it regret} $R(t)$, the difference between the actual reward of an arm-selection rule and the reward of a genie-aided rule that has known statistical information of the arms \cite{Lai_Robbins}. It is proved in \cite{TAC} that for any {\it adaptive allocation rule}\footnote{This means the decisions of the rule are only based on observations in the history \cite{TAC}.} the regret is at least $\mu\ln t$ when $t\rightarrow \infty$, where the factor $\mu$ is determined by the statistical information of arms. A rule that achieves the lower bound of $\mu$ is called {\it efficiently optimal}, and a rule with regret $O(\ln t)$ is called {\it order optimal}. For OSA in cognitive radio networks, reference \cite{LLai} derives order optimal rules to well coordinate the balance between channel exploration and exploitation, with the assumption of perfect channel sensing. Although not efficiently optimal, the rules are {\it sample mean based index rules} \cite{Agrawal}, and their implementation is much simpler than the efficiently optimal rules given in \cite{TAC}. Moreover, a regret bound is also observed with finite $t$ \footnote{In this paper, when we say ``finite $t$", it means sufficiently large and finite $t$.} in rules in \cite{LLai}, while no such bound is observed for finite $t$ in the efficiently optimal rules in \cite{TAC}. A distributed cognitive sensing problem is investigated and formulated as an {\it adversary bandit problem} in \cite{SongG}, where no statistical assumption is made on channel states. Multi-user OSA in distributed manner is investigated in \cite{Animashree_Anandkumar}, modeled as an MABP with multiple players. In the above existing research efforts for OSA in cognitive radio, perfect channel sensing is assumed, and each secondary user can utilize all observed spectrum opportunities (i.e., infinitely backlogged traffic is assumed at the secondary user).

Unlike existing research efforts, this work explores OSA when i) imperfect channel sensing is assumed and ii) a secondary user has only limited ``access demand" (i.e., it may not use all observed spectrum opportunities at a time period). Our motivation for i) is that channel sensing is always imperfect in a real network. And our motivation for ii) is that a user may have only limited traffic to send at a time period (for example, for a voice conversation).\footnote{Actually the case when a secondary user has unlimited access demand can be viewed as a special case of our work.} Similar setup with limited access demand is adopted in \cite{JinJin, Neely, Urgaonkar}. Therefore, unlike existing OSA research where there is only one decision (i.e., to decide which channels to sense, and subsequently access all sensed-free channels), we have two decisions in the OSA in our work: to decide which channels to sense; and if a number of channels are sensed free, to decide which channels to access. Two cases are considered in our work:
\begin{itemize}
\item Case I: when a secondary user can sense all potential channels simultaneously, referred to as {\it full channel sensing};

\item Case II: when a secondary user can sense a subset of the potential channels simultaneously, referred to as {\it partial channel sensing}.

\end{itemize}

Case I is investigate in Section \ref{s:full}, in which we derive OSA rules and theoretically prove that they have asymptotically finite regrets. Case II is investigated in Section \ref{s:partial}, in which we derive OSA rules and theoretically prove that they have regrets $O(\ln t)$ with $t\rightarrow \infty$ and with finite $t$. Performance evaluation of the derived OSA rules is given in Section \ref{s:num}, followed by conclusion remarks in Section \ref{s:con}.

\section{Case I: with Full Channel Sensing}\label{s:full}
Consider a slotted system, where time is partitioned into slots, and the duration of each slot is $T$. For a secondary user, there are $N$ potential primary channels, denoted as Channels $1,2,...,N$, respectively. In each slot, Channel $i$ ($i\in \{1,2,...,N\}$) is free (i.e., without primary activities) with probability $\theta_i$, and $\theta_i$ is unknown by the secondary user. Let $S_i(j)=1$ and $S_i(j)=0$ denote Channel $i$ is free and busy, respectively, at Slot $j$. For each channel, the channel states (busy or free) vary independently from a slot to another. And the $N$ channels have independent channel states.

Each slot consists of a sensing period with duration $\tau$ and data transmission period with duration $T-\tau$. For each slot, during the sensing period the secondary user senses all the $N$ channels. Among all the sensed-free channels, the secondary user can access (i.e., transmit its data over) up to $K$ channels in the data transmission period. For each accessed channel, the transmission rate is denoted $B$.

During the sensing in Slot $j$, denote ${\bf X}(j)=(X_1(j),X_2(j),...,X_N(j))$ as the sensing result of the $N$ channels, where $X_i(j)=1$ and $X_i(j)=0$ mean Channel $i$ is sensed to be free and busy, respectively. Since sensing errors are inevitable, we let $P_d^i$ denote the detection probability of Channel $i$ (i.e., the probability of detecting the primary user activity if there is primary user activity), and $P_f^i$ denote the false-alarm probability of Channel $i$ (i.e., the probability of mistakenly estimating that the primary user is active when there is actually no primary user activity).

Since the secondary user senses all the $N$ channels, the only decision of the secondary user to make is on which channel(s) to access based on its sensing result. To protect primary users, only channels sensed free can be accessed. Since primary users' statistical information $\mathbf{\Theta} \overset{\triangle}=(\theta_1,\theta_2,...,\theta_N)$ is unknown, online learning is needed for the secondary user to estimate $\mathbf{\Theta}$. In the following, we first investigate the situation of single channel access (i.e., $K=1$, the secondary user can or need to access only one channel at a slot), and subsequently extend the research result to the situation of multiple channel access (i.e., $K\ge2$, the secondary user can or need to access more than one channel simultaneously at a slot).

\subsection{Single Channel Access at a Slot ($K=1$)}

To evaluate the performance of a channel access rule, we use the performance of a genie-aided rule (in which the channel statistical information $\mathbf{\Theta}$ is known) as a benchmark for comparison. Until Slot $t$, the expected reward, defined as the total number of bits transmitted by the secondary user, of the genie-aided rule is given as
$\sum\limits_{j=1}^tB(T-\tau)E\left[\mathop {\max}\limits_{i\in\mathcal I(j)}E\left[S_i(j)|X_i(j)=1\right]\right],$
where $\mathcal I(j)$ denotes the set of channels sensed free at Slot $j$, and $E[\cdot]$ denotes expectation. In the reward expression, the outer expectation is for $\mathcal I(j)$, and the inner expectation is for $S_i(j)$.

For any adaptive allocation rule denoted $\psi$, where $\psi(j)=i$ means Channel $i$ is decided to be accessed at Slot $j$, the expected reward until Slot $t$ is
$\sum\limits_{j=1}^tB(T-\tau)\sum\limits_{i=1}^N(1-P_f^i)\theta_i\text{Prob}(\psi(j)=i)$, where $\text{Prob}(\cdot)$ means probability of an event.

The regret (also the learning loss) of rule $\psi$ until Slot $t$, defined as the difference between the expected rewards of $\psi$ and the genie-aided rule, is given as
\begin{equation}\label{e:exp_r_case1}
R(t,\psi)=\sum\limits_{j=1}^tB(T-\tau)E\left[\mathop {\max}\limits_{i\in\mathcal I(j)}E\left[S_i(j)|X_i(j)=1\right]\right]-\sum\limits_{j=1}^tB(T-\tau)\sum\limits_{i=1}^N(1-P_f^i)\theta_i\text{Prob}(\psi(j)=i).
\end{equation}

Since the secondary user can sense all the channels before selecting a channel to access, the channel access process can be modeled as an {\it MABP with side observation} \cite{ChiWang}. For an MABP, it is extremely hard to derive an optimal channel access strategy such that the regret is minimized. Therefore, researchers instead focus on regret bound in asymptotic sense. For example, in \cite{LLai}, asymptotically order optimal rules are derived such that the regret is $O(\ln t)$ when $t\rightarrow \infty$. In our research, we also focus on channel access rule with good asymptotic performance such as asymptotically finite regret. Note that for {\it two-armed} bandit problem with side observation, reference \cite{ChiWang} gives a rule with asymptotically finite regret under {\it direct information} setting. In our work, we derive a rule with asymptotically finite regret for our {\it multi-armed} bandit problem with side observation, as follows.

For sensing of the $N$ channels, we have $2^N$ possible combinations of the sensing result. Denote $\cal{U}$ as the set of the $2^N$ possible combinations. For each $u\in {\cal{U}}$, at each slot (say Slot $t$) we keep a record of $L_u$, which denotes the rate of $u$ as the sensing result, given as the ratio of the number of slots in which $u$ is the sensing result to $t$. Also define $P_u^{{\bf \Theta}^\dag}$ as the probability that $u$ is the sensing result at a slot, which is numerically calculated assuming that ${\bf \Theta}^\dag$ is the vector of free probabilities of the $N$ channels. Our proposed channel access rule is shown in Algorithm \ref{algo1}.


\begin{algorithm}
\caption{Single Channel Access with Full Channel Sensing at Slot $t$}\label{algo1}
\begin{algorithmic}[1]
\STATE Sense $N$ channels, obtain sensing result ${\bf X}(t)$, and update $L_u$, $u \in {\cal U}$.
\STATE Construct candidate set $\mathcal{C}(t)$ of the form
\begin{equation*}
\mathcal{C}(t)= \left\{ {\bf \Theta}^\dag: \sqrt{\sum_{u\in {\cal U}} (P_u^{{\bf \Theta}^\dag} - L_u)^2   }      \le    \mathop {\inf}\limits_{{\bf \Theta}'\in(0,1]^N}  \sqrt{\sum_{u\in {\cal U}} (P_u^{{\bf \Theta}'} - L_u)^2   } +\frac{1}{t}\right\}.
\end{equation*}
\STATE Arbitrarily pick up $\hat{\mathbf{\Theta}} \in \mathcal{C}(t)$, and calculate conditionally expected reward $B(T-\tau)E\left[S_i(t)|X_i(t)=1\right]$ ($i\in\mathcal{I}(t)$) by using $\hat{\mathbf{\Theta}}$ as the vector of channel free probabilities. Here $\mathcal{I}(t)$ denotes the set of channels sensed free at Slot $t$.
\IF{$\mathcal{I}(t)$ is empty}
\STATE Do not access any channel at Slot $t$.
\ELSE
\STATE Access Channel $i^*=\mathop{\arg\max}\limits_{i\in\mathcal{I}(t)}E\left[S_i(t)|X_i(t)=1\right]$.
\ENDIF
\end{algorithmic}
\end{algorithm}

\begin{theory}\label{th:case1_algo1}
Algorithm \ref{algo1} achieves asymptotically finite regret; that is, $\limsup\limits_{t\to\infty}R(t)<\infty$.
\end{theory}
\begin{proof}
See Appendix \ref{a:case1_algo1}.
\end{proof}

Theorem \ref{th:case1_algo1} indicates that the performance of Algorithm \ref{algo1} is surprisingly good through full channel sensing prior to channel access. As a comparison, in the rules derived in \cite{LLai} where the secondary user senses one channel with perfect sensing, performance of $R(t)\sim O(\ln t)$ is achieved, which means the regret goes to infinity when $t\to\infty$.

Algorithm \ref{algo1} suffers from high complexity in the construction of candidate set $\mathcal{C}(t)$ in each slot. To reduce complexity, an alternative channel access rule with linear complexity is introduced, as given in Algorithm \ref{algo2}.
\begin{algorithm}
\caption{Single Channel Access with Full Channel Sensing at Slot $t$}\label{algo2}
\begin{algorithmic}[1]
\STATE Sense $N$ channels, and obtain sensing result $\textbf{X}(t)$.
\STATE Estimate the free probability of Channel $i$ ($i\in \{1,2,...,N\}$) to be $\hat\theta _i(t)=\frac{{\frac{1}{t}\sum\limits_{j=1}^t {X_i(j)}+P_d^i-1}}{{P_d^i-P_f^i}}$.
\STATE Calculate conditionally expected rewards $B(T-\tau)E\left[S_i(t)|X_i(t)=1\right]$, $i\in\mathcal{I}(t)$, by using $\hat{\mathbf{\Theta}}(t)=(\hat\theta_1(t),\hat\theta_2(t),...,\hat\theta_N(t))$ as the vector of channel free probabilities. Here $\mathcal{I}(t)$ denotes the set of channels sensed free at Slot $t$.
\IF{$\mathcal{I}(t)$ is empty}
\STATE Do not access any channel at Slot $t$.
\ELSE
\STATE Access Channel $i^*=\mathop{\arg\max}\limits_{i\in\mathcal{I}(t)}E\left[S_i(t)|X_i(t)=1\right]$.
\ENDIF
\end{algorithmic}
\end{algorithm}

\begin{theory}\label{th:case1_algo2}
Algorithm \ref{algo2} achieves asymptotically finite regret.
\end{theory}
\begin{proof}
See Appendix \ref{a:case1_algo2}.
\end{proof}

\subsection{Multiple Channel Access at a Slot ($K>1$)}

Assume the secondary user can simultaneously access up to $K(>1)$ channels at a slot. Therefore, if the number of channels sensed free at a slot is less than or equal to $K$, then all those sensed-free channels are accessed by the secondary user; otherwise, $K$ channels are selected among the sensed-free channels to be accessed by the secondary user.

We still use the performance of a genie-aided rule with $\bf{\Theta}$ known as a benchmark for comparison. Until Slot $t$, the expected reward of the genie-aided rule is given as
\[\sum\limits_{j=1}^tB(T-\tau)E\Big[\max\limits_{{\cal K}(j)\subset {\cal I}(j),|{\cal K}(j)|\le K}\sum\limits_{i\in{\cal K}(j)}E[S_i(j)|X_i(j)=1]\Big]\]
where $\mathcal I(j)$ denotes the set of channels sensed free at Slot $j$ and $\mathcal K(j)$ denotes the set of channels to be accessed at Slot $j$.

For any adaptive allocation rule $\Psi$ for multiply channel access, where $\Psi(j)$ denotes the set of channels to be accessed at Slot $j$, the expected reward until Slot $t$ is
$\sum\limits_{j=1}^tB(T-\tau)\sum\limits_{i=1}^N(1-P_f^i)\theta_i\text{Prob}(i\in \Psi(j)).$

The regret of rule $\Psi$ is given as $R(t,\Psi)=\sum\limits_{j=1}^tB(T-\tau)E\Big[\max\limits_{{\cal K}(j)\subset {\cal I}(j),|{\cal K}(j)|\le K}\sum\limits_{i\in{\cal K}(j)}E[S_i(j)|X_i(j)=1]\Big]-\sum\limits_{j=1}^tB(T-\tau)\sum\limits_{i=1}^N(1-P_f^i)\theta_i\text{Prob}(i\in\Psi(j))$.

For multiple channel access, we modify Step 7 in Algorithm \ref{algo1} and Algorithm \ref{algo2} as follows: if $|{\cal I}(t)|\le K$, then access all channels in ${\cal I}(t)$; otherwise, among all the channels in ${\cal I}(t)$, access the $K$ channels with the largest $K$ values of $E\left[S_i(t)|X_i(t)=1\right]$. It can be proved that the resulted algorithms have asymptotically finite regret. The proofs are similar to those of Theorems \ref{th:case1_algo1} and \ref{th:case1_algo2}, and are omitted here.

\section{Case II: with Partial Channel Sensing}\label{s:partial}
Still consider $N$ channels. At a slot, the secondary user can sense $M(<N)$ of them and can access up to $K(\le M)$ channels among the sensed-free channels.
Therefore, we have a {\it bi-level MABP}: the first level is to decide which $M$ channels to sense; and the second level is to decide, among the sensed-free channels, which up to $K$ channels to access. The arms played in the two levels are different, which makes the problem much more challenging than classical MABP. To the best of our knowledge, a general bi-level MABP is still an open problem. In the following, we provide solutions to our particular bi-level MABP. Possible extension of our solutions to a more general bi-level MABP is to be investigated in our future work.

Unlike Case I where we have common channel access rules for homogeneous sensing (i.e., $P_d^i=P_d$, $P_f^i=P_f$, $\forall i\in\{1,2,...,N\}$) and heterogeneous sensing (i.e., for each channel, say Channel $i$, we have distinct setting $\{P_d^i,P_f^i\}$), the homogeneous sensing and heterogeneous sensing need to be treated in different ways in Case II, as discussed in Section \ref{s:caseII_homo} and \ref{s:caseII_hete}, respectively.

\subsection{Homogeneous Sensing}\label{s:caseII_homo}

Consider $P_d^i=P_d$, $P_f^i=P_f$, $\forall i\in\{1,2,...,N\}$.  Without loss of generality, we assume $\theta_1>\theta_2>...>\theta_N$.

We still use the performance of a genie-aided rule as a benchmark for comparison. It can be proved that the genie-aided rule should always sense $\mathcal{M}^*=\{1,2,...,M\}$. So until Slot $t$, the expected reward of the genie-aided rule is given as
$U^*(t)=\sum\limits_{j=1}^tE\left[B(T-\tau)\mathop {\max}\limits_{{\cal K}(j) \subset \mathcal{I_{M^*}}(j), |{\cal K}(j)|\le K} \sum\limits_{i\in {\cal K}(j)} E\left[S_i(j)|X_i(j)=1\right]\right]$
where $\mathcal{I_{M^*}}(j)$ denotes the set of sensed-free channels at Slot $j$ if the channels in ${\cal M}^*$ are sensed, and ${\cal K}(j)$ denotes the set of channels to access at Slot $j$.

In the following, we investigate single channel access ($K=1$) and multiple channel access ($K>1$), respectively.

\subsubsection{Single Channel Access at a slot ($K=1$)}

The expected reward of the genie-aided rule until Slot $t$ is:
{\small \begin{equation}
U^*(t)=\sum\limits_{j=1}^tE\left[B(T-\tau)\mathop {\max}\limits_{i\in \mathcal{I_{M^*}}(j)}E\left[S_i(j)|X_i(j)=1\right]\right].
\end{equation}}
Compared with the genie-aided rule, regret of a single channel access rule $\phi$, in which $\phi(j)$ denotes the channel to be accessed at Slot $j$, is given as
\begin{equation}
R(t,\phi)=U^*(t)-\sum\limits_{j=1}^tB(T-\tau)\sum\limits_{i=1}^N(1-P_f^i)\theta_i\text{Prob}(\phi(j)=i).
\end{equation}

Unlike Case I in Section \ref{s:full}, we cannot expect asymptotically finite regret $R(t)$. The reason is as follows. For partial channel sensing, consider a {\it perfect scenario} in which all sensed-free channels are to be accessed and all sensings are perfect. It is shown in Theorem 3.1 in \cite{TAC} and Lemma 2 in \cite{LLai} that the perfect scenario has a lower bound of $O(\ln t)$ on $R(t)$ as $t\to\infty$. It can be proved (the proof is omitted due to space limit) that, if the perfect scenario has regret $C \ln t$ where $C$ is a constant, then our research problem has regret at least $D\ln t$ where $D$ is a constant.

Note that references \cite{TAC} and \cite{Agrawal} give rules with regret $O(\ln t)$ when $t\to\infty$. However, performance of the rules with finite  $t$ is still unclear. In the following, using the UCB1 (here UCB stands for Upper Confidence Bound) in \cite{PAuer}, we derive a channel sensing and access rule that has regret $R(t)\sim O(\ln t)$ with $t\rightarrow \infty$ and with finite $t$. Note that the original UCB1 cannot be directly applied to our research problem, because, if it is directly applied, there is only one decision, i.e., which channels to sense at a slot. Since in our research problem there are two decisions (which channels to sense, and which channel to access among the sensed-free channels), we have necessary extensions to the original UCB1.

At each slot (say Slot $t$), the secondary user keeps records ${\bf T}(t)=(T_1(t), T_2(t),...,T_N(t))$ and ${\bf Y}(t)=(Y_1(t), Y_2(t),...,Y_N(t))$, where $T_i(t)$ is the number of slots in which Channel $i$ has been sensed until Slot $t$, and $Y_i$ is the number of slots in which Channel $i$ has been sensed free until Slot $t$. The proposed channel sensing and access rule is given in Algorithm \ref{algo3}.


\begin{algorithm}
\caption{Single Channel Access with Homogeneous Sensing in Case II (Partial Channel Sensing)}\label{algo3}
\begin{algorithmic}[1]
\STATE Sense all $N$ channels by using $\left\lceil \frac{N}{M} \right\rceil$ slots (where $\lceil \cdot \rceil$ is a ceiling function). At each slot, randomly select one sensed-free channel to access. Update ${\bf T}$ and ${\bf Y}$ at each slot.
\FOR{each subsequent Slot $t$}
\STATE Estimate ${\theta}_i$ ($i=1,2,...,N$) by $\hat{{\theta}}_i(t)= \frac{\frac{Y_i (t-1)}{T_i(t-1)}+P_d-1}{P_d-P_f}$, and determine channel set $\mathcal{M}(t)$ to sense, which includes channels with the $M$ largest indexes $\hat\theta_i(t)+\frac{1}{P_d-P_f}\sqrt {\frac{2\ln (t-1)}{T_i(t-1)}}$.
\STATE Sense channels in $\mathcal{M}(t)$. Let $\mathcal{I}(t)$ denote the set of sensed-free channels. Update $\textbf{T}(t)$ and $\textbf{Y}(t)$.
\IF{$\mathcal{I}(t)$ is nonempty}
\STATE Access Channel $i^*=\mathop{\arg\max}\limits_{i\in\mathcal{I}(t)}\left\{\hat\theta_i(t)+\frac{1}{P_d-P_f}\sqrt {\frac{2\ln (t-1)}{T_i(t-1)}}\right\}$.
\ELSE
\STATE Do not access any channel at Slot $t$.
\ENDIF
\ENDFOR
\end{algorithmic}
\end{algorithm}

\begin{theory}\label{th:case2_single_order_opt}
The regret $R(t)$ of Algorithm \ref{algo3} is $O(\ln t)$ with $t\rightarrow \infty$ and with finite $t$.
\end{theory}
\begin{proof}
See Appendix \ref{a:case2_single_order_opt}.
\end{proof}

\subsubsection{Multiple Channel Access at a slot ($K>1$)}
When the secondary user can simultaneously access $K$ channels at a slot, we modify Algorithm \ref{algo3} as follows: in Step 6, instead of accessing a single channel, the secondary user selects up to $K$ channels in $\mathcal{I}(t)$ with the largest values of $\hat\theta_i(t)+\frac{1}{P_d-P_f}\sqrt {\frac{2\ln (t-1)}{T_i(t-1)}}$. Similar to proof of Theorem \ref{th:case2_single_order_opt}, it can be proved that the regret of the resulted rule is $O(\ln t)$ for finite $t$ and for $t\rightarrow \infty$.

\subsection{Heterogenous Sensing}\label{s:caseII_hete}
 Consider that Channel $i$ ($i=1,...,N$) has distinct setting $\left\{P_d^i,P_f^i\right\}$. The genie-aided rule with known channel statistics $\bf\Theta$ is still used as a benchmark of performance.

When channel statistics $\mathbf{\Theta}$ is unknown, it is desired to find a rule of good performance on regret $R(t)$ under heterogenous sensing. Then a question is raised: can we find a similar rule to those in Section \ref{s:caseII_homo}, with $R(t) \sim O(\ln t)$ for finite $t$ and for $t \rightarrow \infty$? To answer this question, we first look into the insights in the rules in Section \ref{s:caseII_homo}.

As aforementioned, in Case II (partial channel sensing), there are two levels of MABP : the first level is to select which channels to sense, i.e., to select channel set ${\cal M}$ to maximize \[E\left[B(T-\tau)\mathop {\max}\limits_{{\cal K}(j) \subset \mathcal{I_M}(j), |{\cal K}(j)|\le K} \sum\limits_{i\in {\cal K}(j)} E\left[S_i(j)|X_i(j)=1\right]\right]\] while the second level is to select which channels to access, i.e., to select sensed-free  channels with the largest $E\left[S_i(j)|X_i(j)=1\right]$. With homogeneous sensing, the criterion in the first level is simplified to finding the $M$ channels with $M$ largest $\theta_i$'s, while the criterion in the second level is simplified to, among sensed-free channels, finding up to $K$ channels with the largest $\theta_i$'s. Therefore, in Algorithm \ref{algo3}, in both levels we use sample mean of sensing results of each channel, which can be used to estimate $\theta_i$. On the other hand, with heterogeneous sensing, the criteria in the two levels cannot be simplified to finding channels with the largest $\theta_i$'s. Therefore, it is not feasible to use sample mean of sensing results as Algorithm \ref{algo3} does. Rather, we need samples to reflect reward of each arm in each level, as shown in the following.

\subsubsection{Single Channel Access at a Slot ($K=1$)}
Since the secondary user can sense $M$ channels at a slot, the secondary user can sense one from ${N \choose M}$ possible sets of $M$ channels, denoted ${\cal M}_1,{\cal M}_2,...,{\cal M}_{N \choose M}$. In set ${\cal M}_i$ ($i=1,2,...,{N \choose M}$), let $m_{i,j}$ ($j=1,2,...,M$) denote the $j$th channel in ${\cal M}_i$. If the secondary user senses set ${\cal M}_i$ at Slot $t$, let ${\cal I}_{{\cal M}_i}(t)$ represent the sensing result, which is the set of sensed-free channels. Until Slot $t$, let $T_i(t)$ denote the number of time slots in which ${\cal M}_i$ is sensed, and $Y_i(t)$ denote the cumulative reward of the slots in which ${\cal M}_i$ is sensed. Until Slot $t$, let $T_{i,j}(t)$ ($j=1,2,...,M$) denote the number of slots in which ${\cal M}_i$ is sensed and subsequently Channel $m_{i,j}$ is accessed, and $Y_{i,j}(t)$ denote the cumulative reward of Channel $m_{i,j}$ in time slots in which ${\cal M}_i$ is sensed and subsequently Channel $m_{i,j}$ is accessed. Note that when we say ``reward", it means the secondary user transmits over a channel, and receives ACK for the transmission. If no ACK is received, the reward of the corresponding transmission is 0. The proposed channel sensing and access rule is given in Algorithm \ref{algo4}. The secondary user keeps records of $T_i(t)$, $Y_i(t)$, $T_{i,j}(t)$, and $Y_{i,j}(t)$. In the sequel, for simplicity of presentation, the index $(t)$ may be omitted for $T_i(t)$, $Y_i(t)$, $T_{i,j}(t)$, and $Y_{i,j}(t)$.

\begin{algorithm}
\caption{Single Channel Access with Heterogeneous Sensing in Case II (Partial Channel Sensing)}\label{algo4}
\begin{algorithmic}[1]
\FOR{$i=1:{N \choose M}$}
\STATE Keep sensing ${\cal M}_i$ in continuous slots, and at each slot access one free channel that was not accessed before when ${\cal M}_i$ is sensed. This procedure is repeated until each channel in ${\cal M}_i$ has been accessed at least once. For each slot, update $T_i$, $Y_i$, $T_{i,j}$, and $Y_{i,j}$, $j=1,2...,M$.
\ENDFOR
\FOR{each subsequent Slot $t$}
\STATE Calculate indexes $\frac{Y_i}{T_i}+\sqrt{ \frac{2\ln (t-1)}{T_i}}$ ($i\in\{1,2,...,{N\choose M}\}$), and choose $i^\dag=\arg\max\limits_{i=1,...,{N \choose M}} \big\{\frac{Y_i}{T_i}+\sqrt{ \frac{2\ln (t-1)}{T_i}}\big\}$.
\STATE Sense channels in ${\cal M}_{i^\dag}$
\IF{$\mathcal{I}_{{\cal M}_{i^\dag}}(t)$, the set of sensed-free channels at Slot $t$, is nonempty}
\STATE Calculate indexes $\frac{Y_{i^\dag,j}}{T_{i^\dag,j}}+\sqrt{ \frac{2\ln (t-1)}{T_{i^\dag,j}}}$, $m_{i^{\dag},j}\in \mathcal{I}_{{\cal M}_{i^\dag}}(t)$.
\STATE Select $j^\dag=\mathop{\arg\max}\limits_{m_{i^{\dag},j}\in \mathcal{I}_{{\cal M}_{i^\dag}}(t)} \left\{\frac{Y_{i^\dag,j}}{T_{i^\dag,j}}+\sqrt{ \frac{2\ln (t-1)}{T_{i^\dag,j}}}\right\}$, access Channel $m_{i^{\dag},j^{\dag}}$, and check whether the transmission is successful.
\STATE Update $T_{i^\dag}$, $Y_{i^\dag}$, $T_{i^\dag,j^\dag}$, $Y_{i^\dag, j^\dag}$.
\ELSE
\STATE Update $T_{i^\dag}$.
\ENDIF
\ENDFOR
%
\end{algorithmic}
\end{algorithm}

\begin{theory}\label{th:case2_general_order_opt}
The regret $R(t)$ of Algorithm \ref{algo4} is $O(\ln t)$ with $t\rightarrow \infty$ and with finite $t$.
\end{theory}
\begin{proof}
See Appendix \ref{a:case2_general_order_opt}.
\end{proof}

\subsubsection{Multiple Channel Access at a Slot ($K>1$)}

When the secondary user can simultaneously access up to $K$ channels at a slot, we modify Algorithm \ref{algo4} as follows: In Steps 8 and 9, the secondary user selects to access up to $K$ sensed-idle channels with the largest values of $\frac{Y_{i^\dag,j}}{T_{i^\dag,j}}+\sqrt{ \frac{2\ln (t-1)}{T_{i^\dag,j}}}$, $m_{i^\dag, j}\in \mathcal{I}_{{\cal M}_{i^\dag}}(t)$, and updates $T_{i^{\dag},j}$ and $Y_{i^{\dag},j}$ accordingly if Channel $m_{i^\dag,j}$ is accessed. Similarly, it can be proved that the regret of the resulted rule is $O(\ln t)$ with finite $t$ and with $t\rightarrow \infty$.

\section{Performance Evaluation}\label{s:num}
We use Monte-Carlo simulation to validate our analysis. Consider a cognitive radio network with $N=8$ primary channels whose free probabilities are given as $0.9,0.8,0.657,0.564,0.5,$ $0.456,0.404,0.34$ for the 8 channels in our simulation. For homogenous sensing we have $P_d=0.8$ and $P_{f}=0.3$, while in heterogenous sensing we have $(P_d^1,P_d^2,...,P_d^8)=(0.8,0.8,0.7,0.75,0.9,0.67,0.85,$ $0.8)$, and $(P_f^1,P_f^2,...,P_f^8)=(0.3,0.3,0.2,0.25,0.36,0.15,0.32,0.3)$. We also normalize $B(T-\tau)=1$.

\begin{figure}[d]
\begin{center}
\includegraphics[scale=.7]{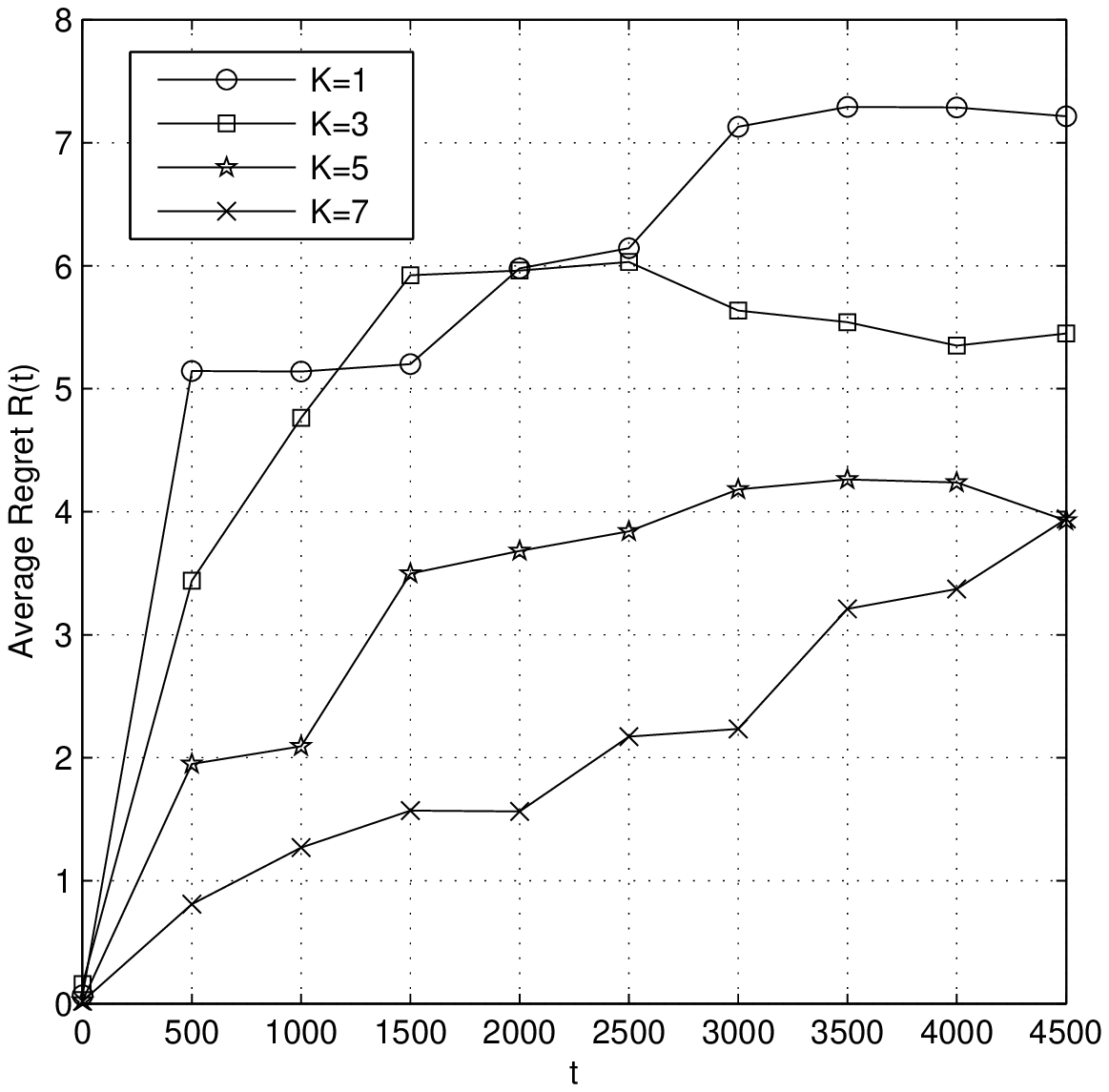}
\caption{Average regret $R(t)$ of Algorithm \ref{algo1} with homogeneous sensing in Case I (full channel sensing)}\label{f:caseI_rule1_homo}
\end{center}
\end{figure}

\begin{figure}[d]
\begin{center}
\includegraphics[scale=.9]{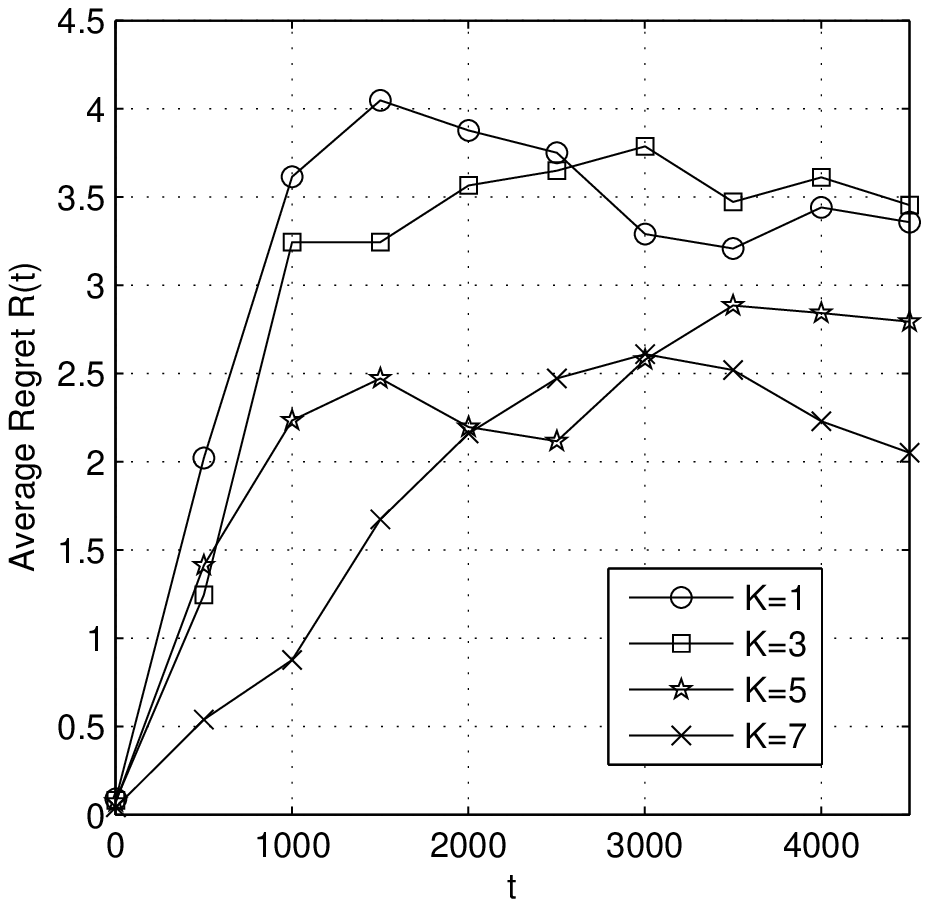}
\caption{Average regret $R(t)$ of Algorithm \ref{algo1} with heterogeneous sensing in Case I (full channel sensing)}\label{f:caseI_rule1_hete}
\end{center}
\end{figure}

\begin{figure}[d]
\begin{center}
\includegraphics[scale=.65]{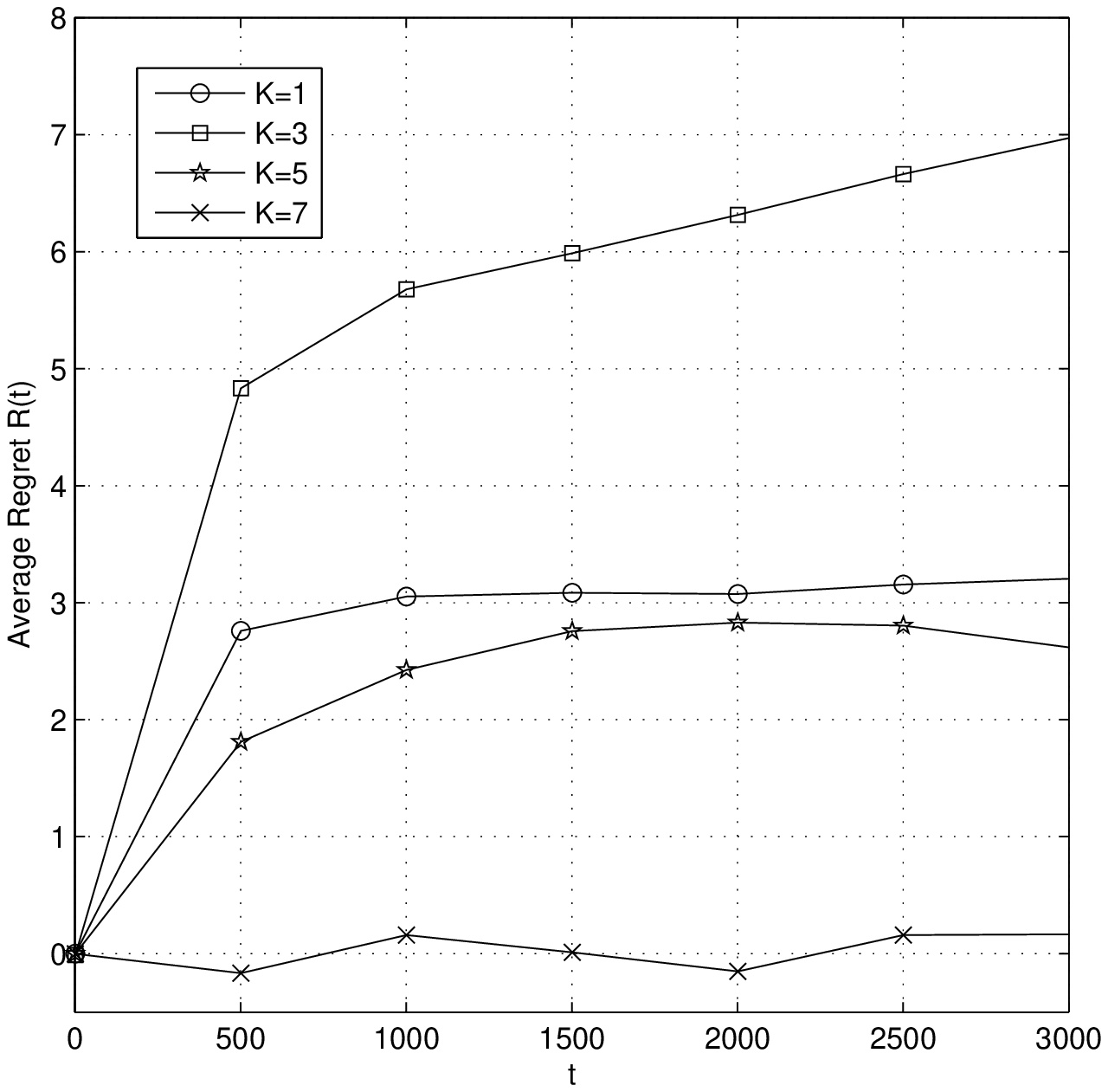}
\caption{Average regret $R(t)$ of Algorithm \ref{algo2} with homogeneous sensing in Case I (full channel sensing)}\label{f:caseI_rule2_homo}
\end{center}
\end{figure}

\begin{figure}[d]
\begin{center}
\includegraphics[scale=.65]{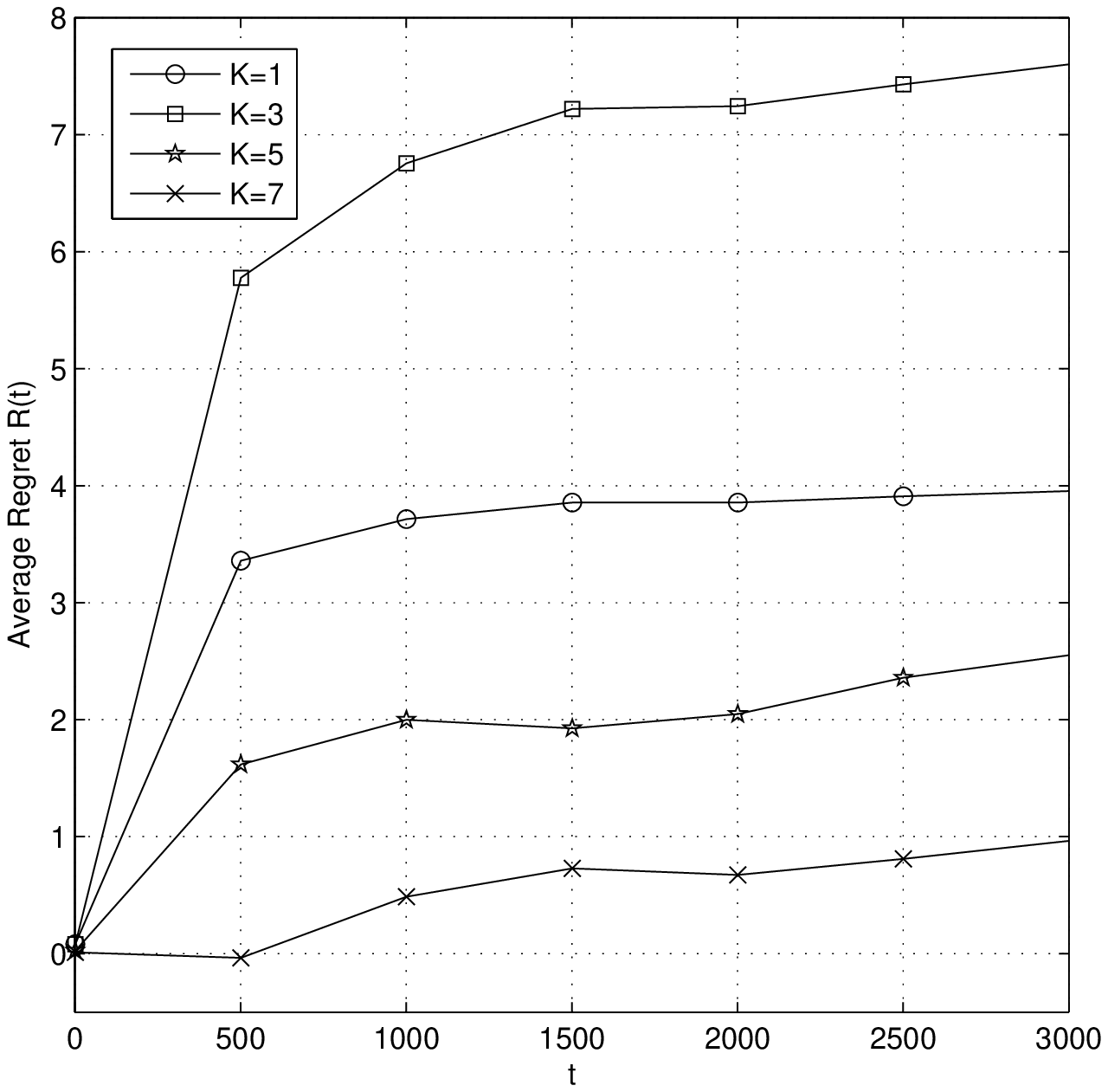}
\caption{Average regret $R(t)$ of Algorithm \ref{algo2} with heterogeneous sensing in Case I (full channel sensing)}\label{f:caseI_rule2_hete}
\end{center}
\end{figure}

Case I with full channel sensing is evaluated first. Figs. \ref{f:caseI_rule1_homo} and \ref{f:caseI_rule1_hete} show the average regret of Algorithm \ref{algo1} with homogeneous sensing and heterogeneous sensing, respectively, while Figs. \ref{f:caseI_rule2_homo} and \ref{f:caseI_rule2_hete} show the average regret of Algorithm \ref{algo2} with homogeneous sensing and heterogeneous sensing, respectively. From the figures it can be seen that when $t$ is large, $R(t)$ tends to be finitely bounded, which is consistent with our analysis in Section \ref{s:full}. Note that, due to complexity of Algorithm \ref{algo1}, Figs. \ref{f:caseI_rule1_homo} and \ref{f:caseI_rule1_hete} are average over only 100 simulation runs, and thus, the regret $R(t)$ does not always increase in the two figures.
Interestingly, in Figs. \ref{f:caseI_rule2_homo} and \ref{f:caseI_rule2_hete}, the $R(t)$ increases when $K$ changes from $1$ to $3$, and $R(t)$ decreases when $K$ further changes to $5$ and $7$. This can be explained as follows. When $K=1$, the false access (i.e., the proposed rule does not access the same channel as the genie-aided rule does) is only on one single channel. When $K$ changes to $3$, the false access is on up to 3 channels, and thus, the reward loss is likely to be larger than that with $K=1$. When $K$ further increases, the up to $K$ channels selected by the proposed rule and the up to $K$ channels selected by the genie-aided rule are likely to be with minor difference, and thus, the reward loss is reduced. When $K=8$ in our example, there is no difference between the channels selected by our proposed rule and the channels selected by the genie-aided rule, which means the reward loss is 0.

\begin{figure}[d]
\begin{center}
\includegraphics[scale=.75]{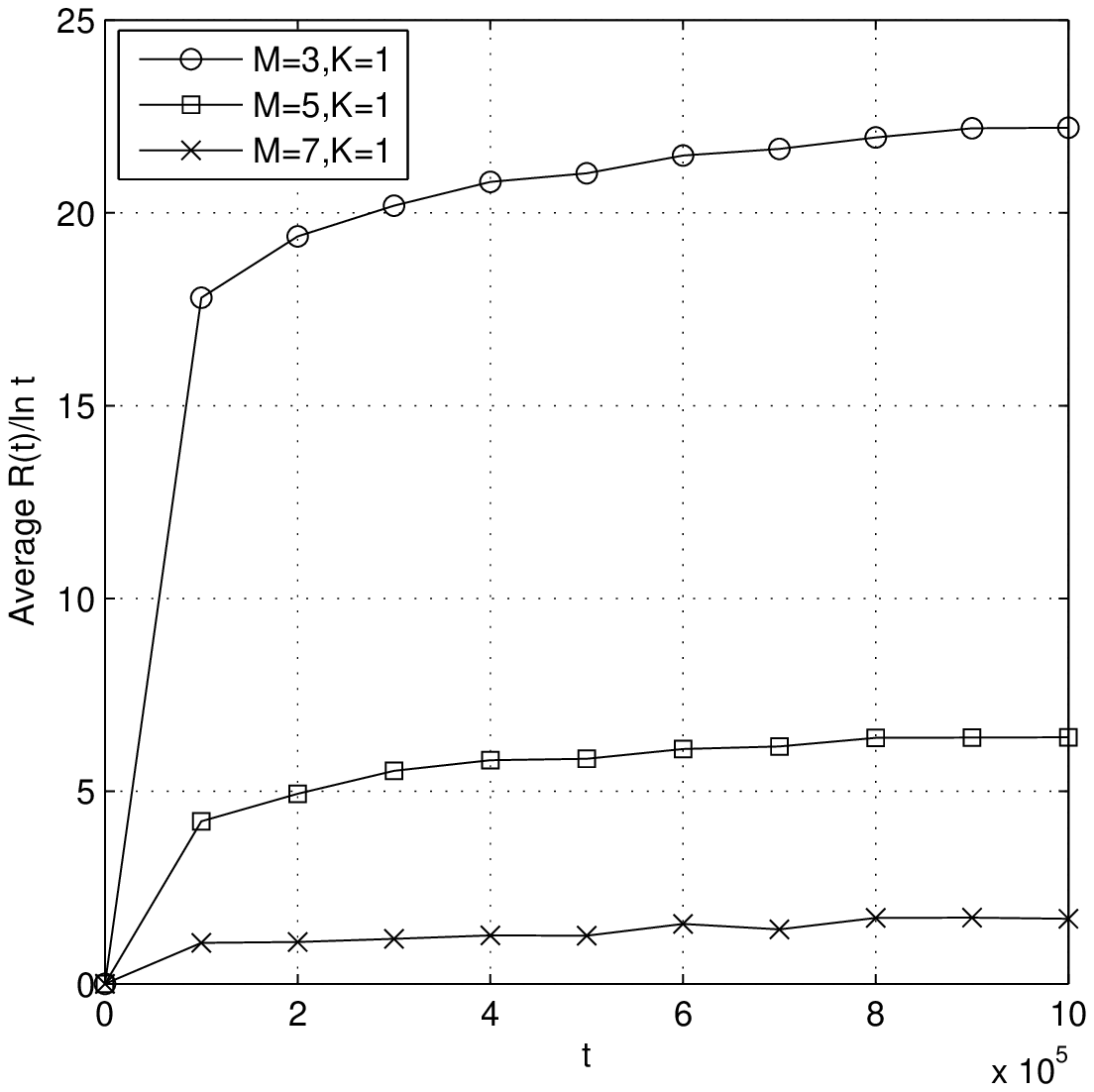}
\caption{Average $R(t)/\ln t$ of Algorithm \ref{algo3} (single channel access) with homogeneous sensing in Case II (partial channel sensing)}\label{f:caseII_homo_single_access}
\end{center}
\end{figure}

\begin{figure}[d]
\begin{center}
\includegraphics[scale=.7]{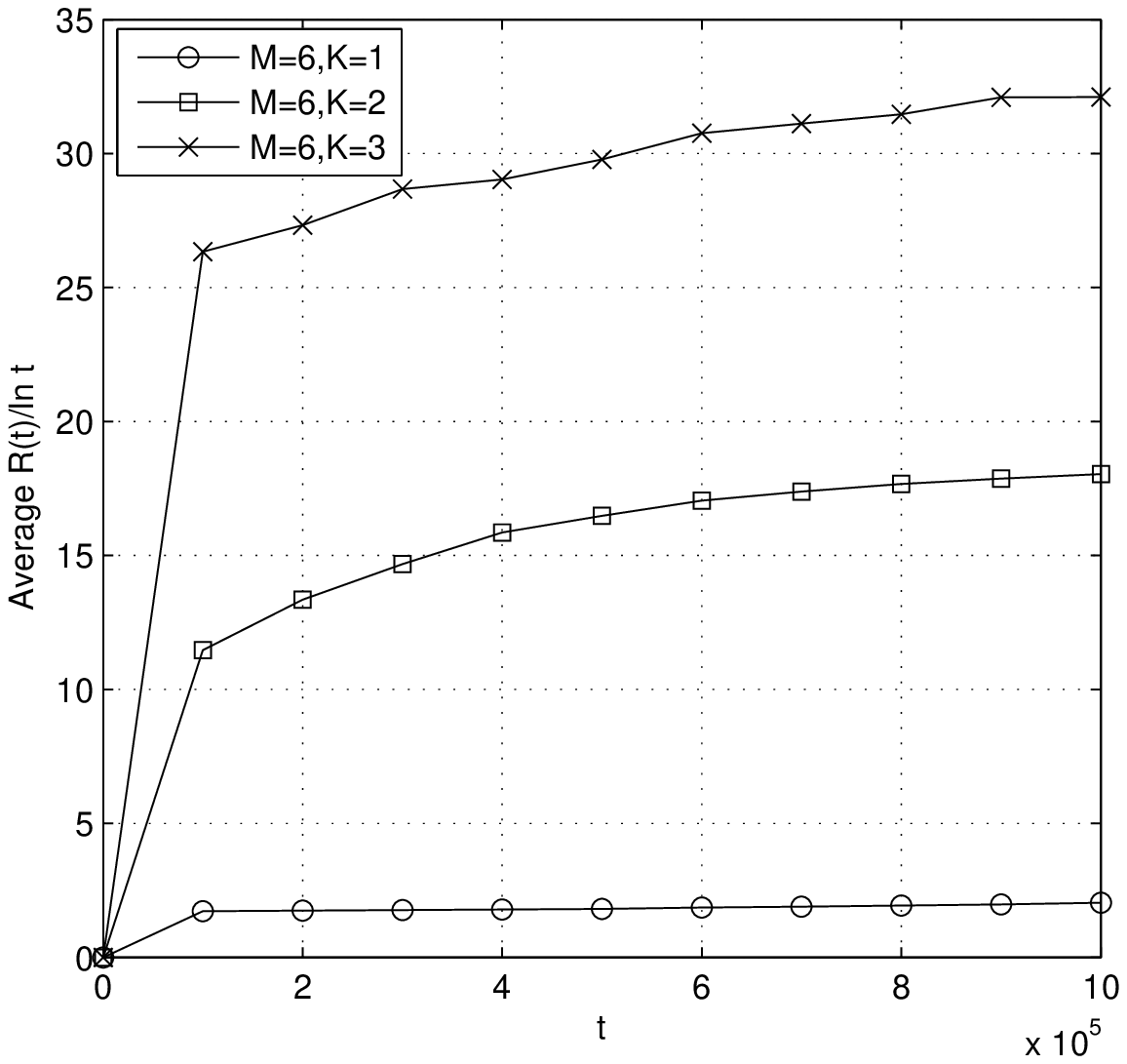}
\caption{Average $R(t)/\ln t$ of proposed multiple channel access rule with homogeneous sensing in Case II (partial channel sensing)}\label{f:caseII_homo_multiple_access}
\end{center}
\end{figure}

\begin{figure}[d]
\begin{center}
\includegraphics[scale=.65]{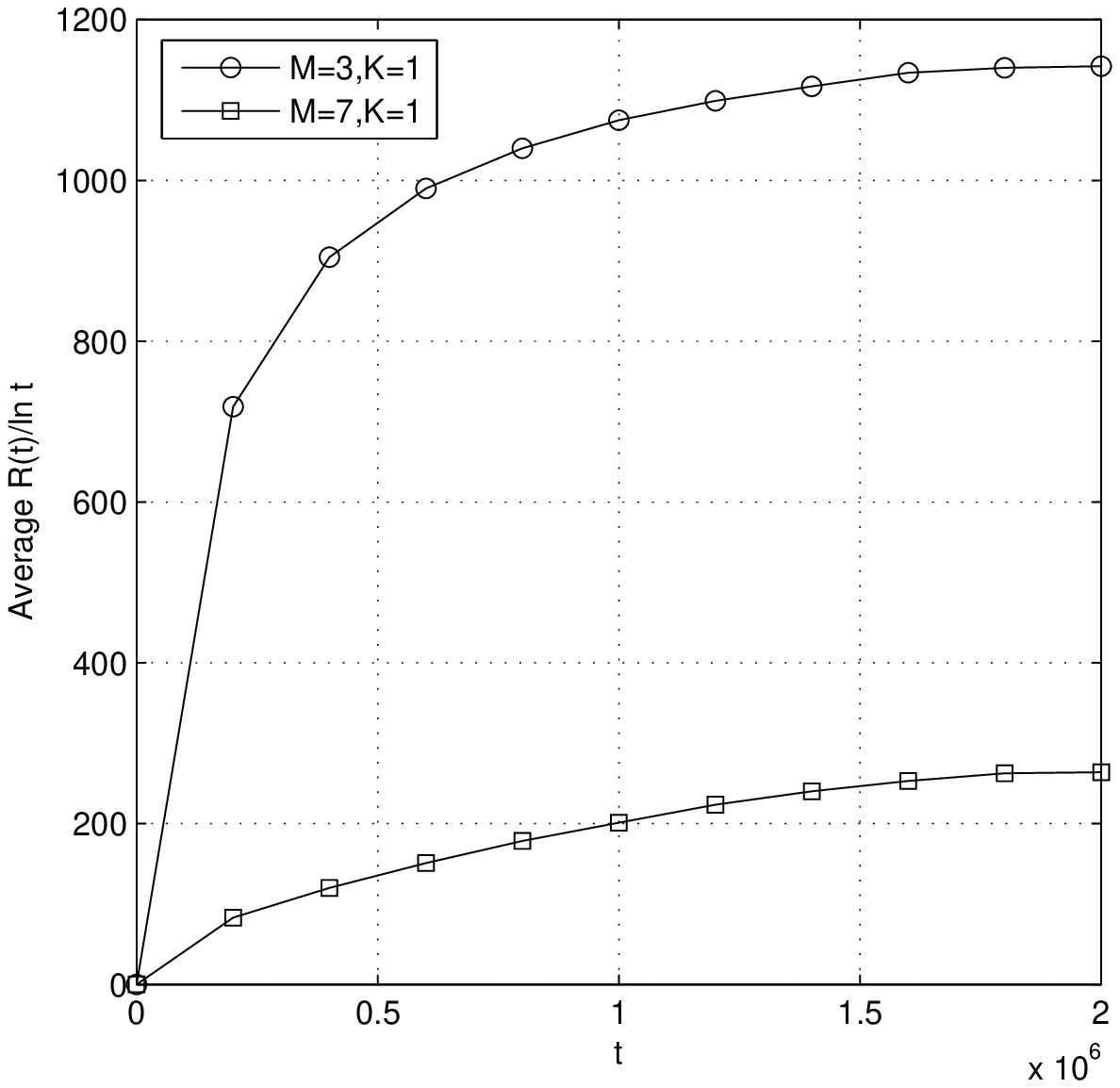}
\caption{Average $R(t)/\ln t$ of Algorithm \ref{algo4} (single channel access) with heterogeneous sensing in Case II (partial channel sensing)}\label{f:caseII_hete_single_access}
\end{center}
\end{figure}

\begin{figure}[d]
\begin{center}
\includegraphics[scale=.65]{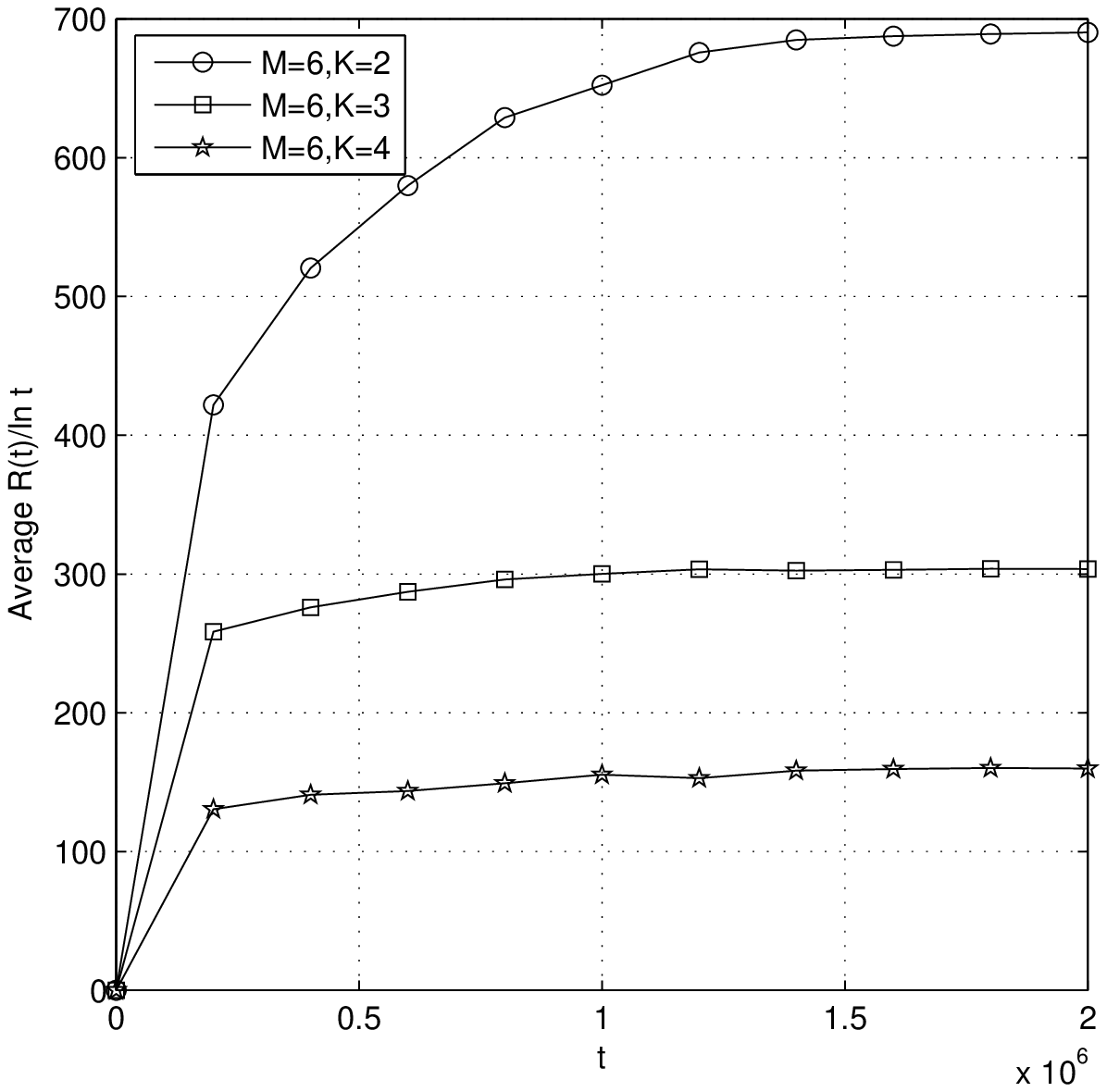}
\caption{Average $R(t)/\ln t$ of proposed multiple channel access rule with heterogeneous sensing in Case II (partial channel sensing)}\label{f:caseII_hete_multiple_access}
\end{center}
\end{figure}

Case II with partial channel sensing is then evaluated. Figs. \ref{f:caseII_homo_single_access} and \ref{f:caseII_homo_multiple_access} show average ${R(t)}/{\ln t}$ in homogeneous sensing with the proposed single channel access and multiple channel access rules, respectively, while Figs. \ref{f:caseII_hete_single_access} and \ref{f:caseII_hete_multiple_access} show average ${R(t)}/{\ln t}$ in heterogeneous sensing with the proposed single channel access and multiple channel access rules, respectively. It can be seen from the four figures that when $t$ is large, average ${R(t)}/{\ln t}$ tends to be finitely bounded, which is consistent with our claim in Section \ref{s:partial} that $R(t)\sim O(\ln t)$.

\section{Conclusion}\label{s:con}
In this paper, the problem of dynamic channel sensing and access by a secondary user in a cognitive radio network is investigated. In the case with full channel sensing, with side information through sensing all the channels, the regret due to unknown primary users' statistical information is proved to be asymptotically finite. On the other hand, for the case with partial channel sensing, asymptotically finite regret cannot be achieved since it is proved that the regret is at least $O(\ln t)$. Therefore, in our research we derive channel sensing and access rules with regret $O(\ln t)$, for homogeneous sensing and heterogeneous sensing, respectively. This research should provide insights to the design of OSA in cognitive radio networks with unknown statistical information of primary channels. Further research may include the case with competition among multiple secondary users and the generalization of our solutions in Case II to a more general bi-level MABP.

\appendices

\section{Proof of Theorem \ref{th:case1_algo1}}\label{a:case1_algo1}
Recall that $\bf \Theta$ is the vector of real channel free probabilities, and in Step 3 of Algorithm \ref{algo1}, $\hat{\bf \Theta}$ is used to estimate $\bf \Theta$. With sensing result ${\bf{X}}(t)$ at Slot $t$, denote $k_{\mathbf{\Theta}}({\bf{X}}(t))$ and $k_{\hat{\mathbf{\Theta}}}({\bf{X}}(t))$ as the best channel which has the largest reward when $\bf \Theta$ and $\hat{\bf \Theta}$ are used as channel availability statistics, respectively.

By following Algorithm \ref{algo1}, the probability of false access (i.e. access a suboptimal channel) is
\begin{equation}\label{e:contract_pe4}
\text{Prob}\left( k_{\hat{\mathbf{\Theta}}} \left({\bf{X}}(t)\right) \ne k_{\mathbf{\Theta}}({\bf{X}}(t))\right) \le \text{Prob}\left(\exists u\in {\cal U}, ~k_{\hat{\mathbf{\Theta}}}(u) \ne k_{\mathbf{\Theta}}(u)\right).
\end{equation}

Define a set $\mathcal{C}_e\triangleq\left\{\mathbf{\Theta}':\exists u\in {\cal U}, ~k_{{\mathbf{\Theta}}'}(u) \ne k_{\mathbf{\Theta}}(u) \right\}$. Then (\ref{e:contract_pe4}) is equivalent to
\begin{equation}\label{e:equivlent_sum}
\text{Prob}\left( k_{\hat{\mathbf{\Theta}}} \left({\bf{X}}(t)\right) \ne k_{\mathbf{\Theta}}({\bf{X}}(t))\right) \le \text{Prob}\left(\hat{\mathbf{\Theta}}\in\mathcal{C}_e\right).
\end{equation}
Define
$\varepsilon \overset{\triangle}= \inf\limits_{\mathbf{\Theta}'\in\mathcal{C}_e}  \sqrt{\sum_{u\in {\cal U}} (P_u^{{\bf \Theta}'}- P_u^{{\bf \Theta}} )^2   }.$
Then we have
$\varepsilon >0$ (the proof for this is omitted due to space limit).

We first consider an event $\left\{ \sqrt{\sum\limits_{u\in {\cal U}} (P_u^{{\bf \Theta}}- L_u)^2} <\frac{\varepsilon }{3}\right\}$ happens. From Algorithm \ref{algo1}, we have
\begin{equation}\label{e:estimate}
\sqrt{\sum\limits_{u\in {\cal U}} (P_u^{\hat{\bf \Theta}}- L_u)^2} \le \mathop {\inf}\limits_{{\bf \Theta}'\in(0,1]^N}  \sqrt{\sum\limits_{u\in {\cal U}} (P_u^{{\bf \Theta}'}- L_u)^2}   +\frac{1}{t}  \le  \sqrt{\sum\limits_{u\in {\cal U}} (P_u^{{\bf \Theta}}- L_u)^2} + \frac{1}{t} < \frac{\varepsilon }{3}+ \frac{1}{t}.
\end{equation}
When $t$ is large enough such that $\frac{1}{t}\le\frac{\varepsilon }{3}$, from (\ref{e:estimate}) we have
\begin{equation}\label{e:derive_in_th1}
\sqrt{\sum_{u\in {\cal U}} (P_u^{{\bf \Theta}}- P_u^{\hat{\bf \Theta}} )^2   }    \le \sqrt{\sum_{u\in {\cal U}} (P_u^{{\bf \Theta}}- L_u)^2} + \sqrt{\sum_{u\in {\cal U}} (P_u^{\hat{\bf \Theta}}- L_u)^2} <\varepsilon
\end{equation}
which means $\hat{\mathbf{\Theta}}(t)\notin \mathcal{C}_e$ from the definition of $\varepsilon$. It also means that, if  $\hat{\mathbf{\Theta}}(t)\in \mathcal{C}_e$, then we should have $ \sqrt{\sum\limits_{u\in {\cal U}} (P_u^{{\bf \Theta}}- L_u)^2} \ge \frac{\varepsilon }{3}$. Then we have
\begin{equation}\label{e:contract_pe}
\text{Prob}\left(\hat{\mathbf{\Theta}}\in \mathcal{C}_e\right)\le \text{Prob}\left(\sqrt{\sum\limits_{u\in {\cal U}} (P_u^{{\bf \Theta}}- L_u)^2} \ge \frac{\varepsilon }{3}\right)   \le a(t)\overset{\triangle}= (t+1)^{2^N} e^{-t\inf\limits_{\{L_u'\}_{u\in {\cal U}}\in \mathcal{B}} \sum\limits_{u\in{\cal U}}L'_u\ln (L'_u/P_u^{\bf\Theta})
}
\end{equation}
where the second inequality comes from the Sanov Theorem (i.e., Theorem 2.1.10) in \cite{Dembo}, and $\mathcal{B}$ denotes a vector space $\left\{\{L'_u\}_{u\in{\cal U}}:\sqrt{\sum\limits_{u\in{\cal U}}(P_u^{\bf\Theta}-L'_u)^2}
\ge\frac{\varepsilon}{3}\right\}$, which is closed.

For the exponent in the expression of $a(t)$, we have
\begin{equation}\label{e:jessen}
\sum\limits_{u\in{\cal U}}L'_u\ln (L'_u/P_u^{\bf\Theta})=\sum\limits_{u\in{\cal U}}\left(P_u^{\bf\Theta}\frac{L'_u}{P_u^{\bf\Theta}}\ln (L'_u/P_u^{\bf\Theta})\right)
\ge \left(\sum\limits_{u\in{\cal U}}P_u^{\bf\Theta}\frac{L'_u}{P_u^{\bf\Theta}}\right)\ln \left({\sum\limits_{u\in{\cal U}}P_u^{\bf\Theta}\frac{L'_u}{P_u^{\bf\Theta}}}\right)=0
\end{equation}
where the inequality comes from the Jensen's inequality and the fact that $x\ln x$ is a convex function. In addition, $\sum\limits_{u\in{\cal U}}L'_u\ln (L'_u/P_u^{\bf\Theta})$ is continuous and strictly convex, which, together with $\varepsilon>0$ and (\ref{e:jessen}), leads to $\inf\limits_{\{L'_u\}_{u\in{\cal U}}\in \mathcal{B}}\sum\limits_{u\in{\cal U}}L'_u\ln (L'_u/P_u^{\bf\Theta})>0$. And thus, from the definition of $a(t)$ given in (\ref{e:contract_pe}), we have $\lim\limits_{t\to \infty} \frac{a({t+1})}{a({t})}<1$.

From (\ref{e:equivlent_sum}) and (\ref{e:contract_pe}), we have
$ \text{Prob}\left( k_{\hat{\mathbf{\Theta}}} \left({\bf{X}}(t)\right) \ne k_{\mathbf{\Theta}}({\bf{X}}(t))\right) \le a(t)$ when $\frac{1}{t}\le\frac{\varepsilon }{3}$.
So for regret $R(t)$ of Algorithm \ref{algo1}, we have
\begin{align}
\limsup\limits_{t\to\infty}R(t)
&\le c_0 \sum\limits_{j=1}^{\lfloor\frac{3}{\varepsilon}\rfloor} \text{Prob}\left( k_{\hat{\mathbf{\Theta}}} \left({\bf{X}}(j)\right)\ne k_{\mathbf{\Theta}}({\bf{X}}(j))\right)+ c_0 \lim\limits_{t \to \infty } \sum\limits_{j=\lfloor\frac{3}{\varepsilon}\rfloor+1}^t \text{Prob}\left( k_{\hat{\mathbf{\Theta}}} \left({\bf{X}}(j)\right) \ne k_{\mathbf{\Theta}}({\bf{X}}(j))\right)\nonumber
\\&
\le c_0\left\lfloor\frac{3}{\varepsilon}\right\rfloor + c_0 \lim\limits_{t \to \infty } \sum\limits_{j=\lfloor\frac{3}{\varepsilon}\rfloor+1}^t a(j)<\infty\label{e:finite_algo1}
\end{align}
where $\lfloor \cdot \rfloor$ is a floor function, $c_0$ denotes the largest possible reward loss due to false access in a slot, which is finite, and the last inequality comes from $\lim\limits_{t\to \infty} \frac{a({t+1})}{a({t})}<1$.

Therefore, by following Algorithm \ref{algo1}, asymptotically finite regret is achieved.

\section{Proof of Theorem \ref{th:case1_algo2}}\label{a:case1_algo2}
For Algorithm \ref{algo2}, the probability of false access is calculated as
\begin{equation}\label{e:algo2_false_prob}
\text{Prob}\left( k_{\hat{\mathbf{\Theta}}(t)} \left(\textbf{X}(t)\right)\ne k_{\mathbf{\Theta}}\left(\textbf{X}(t)\right)\right)
=\sum\limits_{u\in {\cal U}} \text{Prob}\left(k_{\hat{\mathbf{\Theta}}(t)}(u)\ne k_{\mathbf{\Theta}}(u)\right)\text{Prob}\left(\textbf{X}(t)=u\right)
\end{equation}

in which
\begin{align}
&\text{Prob}\left(k_{\hat{\mathbf{\Theta}}(t)}(u)\ne k_{\mathbf{\Theta}}(u)\right)
=\text{Prob}\left(\mathop{\arg\max}_{i\in \mathcal{I}_u} \frac{(1-P_f^i)\hat{\theta _i}(t)}{f\left(\hat{\theta_i}(t)\right)}\ne\mathop{\arg\max}_{i\in \mathcal{I}_u}\frac{(1-P_f^i)\theta_i }{f(\theta_i)} \right)\nonumber
\\&\le \sum\limits_{i>k, i\in \mathcal{I}_u, k\in \mathcal{I}_u} \text{Prob}\left(\frac{(1-P_f^{\pi(i)})\hat\theta _{\pi(i)}(t)}{f\left(\hat\theta_{\pi(i)}(t)\right)}>\frac{(1-P_f^{\pi(k)})\hat\theta _{\pi(k)}(t)}{f\left(\hat\theta_{\pi(k)}(t)\right)}\right)\label{e:err_prob111}
\end{align}
where $\mathcal{I}_u$ is the set of sensed-free channels when the sensing result is $\textbf{X}(t)=u$,
${f(\theta_i)}=(1-P_{f}^i)\theta_i+(1-P_d^i)(1-\theta_i),$
and $(\pi(1), \pi(2), ...,\pi(N))$ is a permutation of $(1,2,...,N)$ such that $\frac{(1-P_f^{\pi(1)})\theta_{\pi(1)} }{f(\theta_{\pi(1)})}>\frac{(1-P_f^{\pi(2)})\theta_{\pi(2)} }{f(\theta_{\pi(2)})}>...>\frac{(1-P_f^{\pi(N)})\theta_{\pi(N)} }{f(\theta_{\pi(N)})}$.

First consider homogeneous sensing when $P_d^i=P_d$ and $P_f^i=P_f$, $i\in \{1,2,...,N\}$. Without loss of generality, assume $\{\theta_1>\theta_2>...>\theta_N\}$. Then (\ref{e:err_prob111}) is simplified as
\begin{equation}\label{e:err_prob111_sim}
\text{Prob}\left(k_{\hat{\mathbf{\Theta}}(t)}(u)\ne k_{\mathbf{\Theta}}(u)\right) \le \sum\limits_{i>k, i\in \mathcal{I}_u, k\in \mathcal{I}_u} \text{Prob}\left(\hat{\theta}_i(t)>\hat {\theta}_k(t)\right).
\end{equation}

According to Algorithm \ref{algo2}, we have $\hat{\theta }_i(t)=\frac{{\frac{1}{t}\sum\limits_{j =1}^t X_i(j)+P_d-1}}{{P_d-P_f}}$ to estimate $\theta_i$. We denote the sum of sensing samples until Slot $t$ for Channel $i$ ($i=1,2,...,N$) as $X_i^t\triangleq\sum\limits_{j=1}^tX_i(j)$. So $X_1^t, X_2^t,...,X_N^t$ are independent binomial random variables with parameters $f(\theta_1), f(\theta_2), ..., f(\theta_N)$, respectively. When $t$ is large enough (say $t\ge t_0$), the binomial distribution of $X_i^t$ can be approximated as a normal distribution with mean $tf(\theta_i)$ and variance $tf(\theta_i)(1-$ $f(\theta_i))$. We use $g_{X_i^t}$ to denote the probability density function of ${X_i^t}$, which follows a normal distribution. Then for the term in the summation in (\ref{e:err_prob111_sim}), we have
\begin{align}
&\text{Prob}\left(\hat{\theta}_i(t)>\hat {\theta }_k(t)\right)=\text{Prob}\left(X_i^t>X_k^t\right)
=\int_{-\infty}^{+\infty} g_{X_k^t}(y)\int_y^{+\infty} g_{X_i^t}(x)\,\mathrm{d}x\mathrm{d}y
\nonumber
\\&=\int_{-\infty}^{tf(\theta_i)} g_{X_k^t}(y)\int_y^{+\infty} g_{X_i^t}(x)\,\mathrm{d}x\mathrm{d}y +\int_{tf(\theta_i)}^{+\infty} g_{X_k^t}(y)\int_y^{+\infty} g_{X_i^t}(x)\,\mathrm{d}x\mathrm{d}y.  \label{e:int_dec}
\end{align}
The two terms on the right hand side of (\ref{e:int_dec}) have the following upper bounds.
{\small \begin{equation}
\int_{-\infty}^{tf(\theta_i)} g_{X_k^t}(y)\int_y^{+\infty} g_{X_i^t}(x)\,\mathrm{d}x\mathrm{d}y<\int_{-\infty}^{tf(\theta_i)} g_{X_k^t}(y)\mathrm{d}y=Q\left(\frac {t\left(f(\theta_k)-f(\theta_i)\right)}{\sqrt {tf(\theta_k)\left(1-f(\theta_k)\right)}}\right)
\le \frac{1}{2}e^{-\frac {\left(f(\theta_k)-f(\theta_i)\right)^2}{2f(\theta_k)\left(1-f(\theta_k)\right)}t}
\end{equation}}
where the second inequality comes from the Chernoff bound. Here $Q(\cdot)$ is the $Q$-function given as $Q(x)=\frac{1}{\sqrt{2\pi}}\int_{x}^{\infty} e^{-\frac{v^2}{2}} dv $.
\begin{align}
&\int_{tf(\theta_i)}^{+\infty}g_{X_k^t}(y)\int_y^{+\infty} g_{X_i^t}(x)\,\mathrm{d}x\mathrm{d}y \nonumber
\\&
\le \int_{tf(\theta_i)}^{+\infty}\frac{1}{2\sqrt {2\pi} \sqrt {f(\theta_k)(1-f(\theta_k))t}}e^{-\frac{(y-f(\theta_k) t)^2}{2f(\theta_k)(1-f(\theta_k))t}}e^{-\frac{(y-f(\theta_i)t)^2}{2f(\theta_i)(1-f(\theta_i))t}}\mathrm{d}y
\nonumber
\\&
\overset{\substack{R_i=f(\theta_i)(1-f(\theta_i))t\\R_k=f(\theta_k)(1-f(\theta_k))t}}{=}\frac {1}{2}\int_{tf(\theta_i)}^{+\infty} \frac{1}{\sqrt {2\pi f(\theta_k)(1-f(\theta_k))t}}e^{-\frac{R_i(y-f(\theta_k)t)^2+R_k(y-f(\theta_i)t)^2}{2R_iR_k}}\mathrm{d}y
\nonumber\\&=\frac {1}{2}\frac{1}{\sqrt {R_k}}\sqrt{\frac{R_kR_i}{R_i+R_k}} e^{-\frac{\frac{R_i(f(\theta_k)t)^2+R_k(f(\theta_i)t)^2}{R_i+R_k}-\frac{\left(R_if(\theta_k)+R_kf(\theta_i)\right)^2t^2}{(R_i+R_k)^2}}{\frac{2R_iR_k}{R_i+R_k}}}
Q\left(\frac{tf(\theta_i)-t\frac{R_if(\theta_k)+R_kf(\theta_i)}{R_i+R_k}}{\sqrt{\frac{R_iR_k}{R_i+R_k}}}\right)
\nonumber\\&
\le \frac{1}{4}\sqrt{\frac{R_i}{R_i+R_k}} e^{-\frac{(\theta_i-\theta_k)^2t^2}{2(R_i+R_k)}}
=\frac{1}{4}\sqrt{\frac{R_i}{R_i+R_k}} e^{-\frac{(\theta_i-\theta_k)^2}{2\left(f(\theta_i)(1-f(\theta_i))
+f(\theta_k)(1-f(\theta_k))\right)}t}\label{e:int2}
\end{align}
where the two inequalities are from the Chernoff bound.

From (\ref{e:algo2_false_prob}) and (\ref{e:err_prob111_sim})-(\ref{e:int2}), we can bound the false access probability, for Slot $t$ when $t\ge t_0$, as
{\small \begin{align*}
&\text{Prob}\left( k_{\hat{\mathbf{\Theta}}(t)} \left(\textbf{X}(t)\right)\ne k_{\mathbf{\Theta}}\left(\textbf{X}(t)\right)\right)\le
\sum\limits_{u\in {\cal U}}\sum\limits_{i>k, i\in {\cal I}_u, k\in {\cal I}_u} \text{Prob}\left(\hat{\theta}_i(t)>\hat {\theta }_k(t)\right)
\text{Prob}\left(\textbf{X}(t)=u\right)\\&
\le \sum\limits_{u\in {\cal U}} \sum\limits_{i>k, i\in {\cal I}_u, k\in {\cal I}_u} \left(\frac{1}{2}e^{-\frac {\left(f(\theta_k)-f(\theta_i)\right)^2}{2f(\theta_k)\left(1-f(\theta_k)\right)}t}+
\frac{1}{4}\sqrt{\frac{R_i}{R_i+R_k}} e^{-\frac{(\theta_i-\theta_k)^2}{2(f(\theta_i)(1-f(\theta_i))
+f(\theta_k)(1-f(\theta_k)))}t}\right)\text{Prob}\left(\textbf{X}(t)=u\right)
\\&
\le c_1e^{-c_2t}, \text{~where } c_1={|{\cal U} | \choose 2},c_2=\min\limits_{i>k}\left\{\frac {\left(f(\theta_k)-f(\theta_i)\right)^2}{2f(\theta_k)\left(1-f(\theta_k)\right)}, \frac{(\theta_i-\theta_k)^2}{2(f(\theta_i)(1-f(\theta_i))
+f(\theta_k)(1-f(\theta_k)))}\right\}.
\end{align*}}

Then for regret $R(t)$ of Algorithm \ref{algo2}, we have
\begin{align*}
&\limsup\limits_{t\to\infty}R(t)
\le \limsup\limits_{t\to\infty}\sum\limits_{j=1}^t c_0\text{Prob}\left( k_{\hat{\mathbf{\Theta}}(j)} \left(\textbf{X}(j)\right)\ne k_{\mathbf{\Theta}}\left(\textbf{X}(j)\right)\right)
\\&
\le \sum\limits_{j=1}^{t_0} c_0\text{Prob}\left( k_{\hat{\mathbf{\Theta}}(j)} \left(\textbf{X}(j)\right)\ne k_{\mathbf{\Theta}}\left(\textbf{X}(j)\right)\right)+
\limsup\limits_{t\to\infty}\sum\limits_{j=t_0+1}^t c_0\text{Prob}\left( k_{\hat{\mathbf{\Theta}}(j)} \left(\textbf{X}(j)\right)\ne k_{\mathbf{\Theta}}\left(\textbf{X}(j)\right)\right)
\\&
\le c_0t_0+\sum\limits_{j=t_0+1}^\infty c_0c_1e^{-c_2j}<\infty
\end{align*}
where $c_0$ denotes the largest possible reward loss by accessing a false channel (i.e., a suboptimal channel) at a slot.

Then consider heterogenous sensing when we do not have $P_d^i=P_d$ and $P_f^i=P_f$, $i\in \{1,2,...,N\}$. Without loss of generality, assume $\frac{(1-P_f^1)\theta_1 }{f(\theta_1)}>...>\frac{(1-P_f^N)\theta_N }{f(\theta_N)}$. Then (\ref{e:err_prob111}) is rewritten as
\begin{equation}\label{e:single_het_prob}
\begin{array}{lll}
&\text{Prob}\left(k_{\hat{\mathbf{\Theta}}(t)}(u)\ne k_{\mathbf{\Theta}}(u)\right) \le \sum\limits_{i>k, i\in \mathcal{I}_u, k\in \mathcal{I}_u} \text{Prob}\left(\frac{(1-P_{f}^i)\hat\theta _i(t)}{f(\hat\theta_i(t))}>\frac{(1-P_{f}^k)\hat\theta _k(t)}{f(\hat\theta_k(t))}\right)\\
&=\sum\limits_{i>k, i\in \mathcal{I}_u, k\in \mathcal{I}_u}  \text{Prob}\left(X_i^tX_k^t\left(
\frac{1-P_f^i}{P_d^i-P_f^i}-\frac{1-P_f^k}{P_d^k-P_f^k}\right)>
t\left(\frac{(1-P_f^i)(1-P_d^i)}{P_d^i-P_f^i}X_k^t-\frac{(1-P_f^k)(1-P_d^k)}{P_d^k-P_f^k}X_i^t\right)
\right)
\end{array}
\end{equation}
where the second line comes from $\hat\theta _i(t)=\frac{{\frac{1}{t}\sum\limits_{j=1}^t {X_i(j)}+P_d^i-1}}{{P_d^i-P_f^i}}=\frac{{\frac{1}{t}X_i^t+P_d^i-1}}{{P_d^i-P_f^i}}$.

Define $d_1\triangleq\frac{(1-P_f^i)(1-P_d^i)}{(P_d^i-P_f^i)}\Big/\left(\frac{1-P_f^i}{P_d^i-P_f^i}-\frac{1-P_f^k}{P_d^k-P_f^k}\right)$ and $d_2\triangleq\frac{(1-P_f^k)(1-P_d^k)}{(P_d^k-P_f^k)}\Big/\left(\frac{1-P_f^i}{P_d^i-P_f^i}-\frac{1-P_f^k}{P_d^k-P_f^k}\right)$. Then (\ref{e:single_het_prob}) can be rewritten as
\begin{equation}\label{e:single_het_prob_rewrite}
\begin{array}{lll}
&\text{Prob}\left(k_{\hat{\mathbf{\Theta}}(t)}(u)\ne k_{\mathbf{\Theta}}(u)\right) \le \sum\limits_{i>k, i\in \mathcal{I}_u, k\in \mathcal{I}_u}  \text{Prob}\left(X_i^tX_k^t>td_1X_k^t-td_2X_i^t\right)
\\&
= \sum\limits_{i>k, i\in \mathcal{I}_u, k\in \mathcal{I}_u}  \left(\int_{td_1}^{+\infty}g_{X_i^t}(x)\int_{\frac{td_2x}{td_1-x}}^{+\infty} g_{X_k^t}(y)\,\mathrm{d}y\mathrm{d}x+\int_{-\infty}^{td_1} g_{X_i^t}(x)\int_{-\infty}^{\frac{td_2x}{td_1-x}} g_{X_k^t}(y)\,\mathrm{d}y\mathrm{d}x\right).
\end{array}
\end{equation}

In order to get a bound of $\text{Prob}\left(k_{\hat{\mathbf{\Theta}}(t)}(u)\ne k_{\mathbf{\Theta}}(u)\right)$, next we derive the bounds for the two terms in the summation in the last line in (\ref{e:single_het_prob_rewrite}). Without loss of generality, we assume $\frac{1-P_f^i}{P_d^i-P_f^i}-\frac{1-P_f^k}{P_d^k-P_f^k}>0$, while scenario with $\frac{1-P_f^i}{P_d^i-P_f^i}-\frac{1-P_f^k}{P_d^k-P_f^k}<0$ can be similarly proved. Note that when $\frac{1-P_f^i}{P_d^i-P_f^i}-\frac{1-P_f^k}{P_d^k-P_f^k}=0$, similar way to that in the homogenous sensing can be used to derive a bound of $\text{Prob}\left(k_{\hat{\mathbf{\Theta}}(t)}(u)\ne k_{\mathbf{\Theta}}(u)\right)$.
\begin{equation*}
\int_{td_1}^{+\infty} g_{X_i^t}(x)\int_{\frac{td_2x}{td_1-x}}^{+\infty} g_{X_k^t}(y)\,\mathrm{d}y\mathrm{d}x
\le\int_{td_1}^{+\infty} g_{X_i^t}(x)\mathrm{d}x=Q\left(\frac{td_1-tf(\theta_i)}{\sqrt{f(\theta_i)(1-f(\theta_i))t}}\right)\le\frac{1}{2}e^{-\frac{(d_1-f(\theta_i))^2}
{2f(\theta_i)(1-f(\theta_i))}t}
\end{equation*}
where the last inequality comes from the Chernoff bound, in which the following fact is used:
\begin{equation*}
d_1=\frac{(1-P_f^i)(1-P_d^i)}{(P_d^i-P_f^i)}\Big/\left(\frac{1-P_f^i}{P_d^i-P_f^i}-\frac{1-P_f^k}{P_d^k-P_f^k}\right)>\frac{(1-P_f^i)(1-P_d^i)}{1-P_f^i-(P_d^i-P_f^i)}=1-P_f^i\ge f(\theta_i).
\end{equation*}
The second term in the summation in the last line in (\ref{e:single_het_prob_rewrite}) is decomposed into two sub-terms:
\begin{align}
&\int_{-\infty}^{td_1} g_{X_i^t}(x)\int_{-\infty}^{\frac{td_2x}{td_1-x}} g_{X_k^t}(y)\,\mathrm{d}y\mathrm{d}x\nonumber
\\&
=\int_{\frac{tf(\theta_k)d_1}{d_2+f(\theta_k)}}^{td_1} g_{X_i^t}(x)\int_{-\infty}^{\frac{td_2x}{td_1-x}} g_{X_k^t}(y)\,\mathrm{d}y\mathrm{d}x+
\int_{-\infty}^{\frac{tf(\theta_k)d_1}{d_2+f(\theta_k)}} g_{X_i^t}(x)\int_{-\infty}^{\frac{td_2x}{td_1-x}} g_{X_k^t}(y)\,\mathrm{d}y\mathrm{d}x.\label{e:second_term_decom}
\end{align}
The first sub-term in (\ref{e:second_term_decom}) is bounded as
\begin{align}
&\int_{\frac{tf(\theta_k)d_1}{d_2+f(\theta_k)}}^{td_1} g_{X_i^t}(x)\int_{-\infty}^{\frac{td_2x}{td_1-x}} g_{X_k^t}(y)\,\mathrm{d}y\mathrm{d}x<\int_{\frac{tf(\theta_k)d_1}{d_2+f(\theta_k)}}^{td_1} g_{X_i^t}(x)\mathrm{d}x<\int_{\frac{tf(\theta_k)d_1}{d_2+f(\theta_k)}}^{+\infty} g_{X_i^t}(x)\mathrm{d}x\nonumber
\\&=Q\left(\frac{t\left(\frac{f(\theta_k)d_1}{d_2+f(\theta_k)}-f(\theta_i)\right)}{\sqrt{f(\theta_i)(1-f(\theta_i))t}}\right)
\le\frac{1}{2}e^{-\frac{\left(\frac{f(\theta_k)d_1}{d_2+f(\theta_k)}-f(\theta_i)\right)^2}{2f(\theta_i)(1-f(\theta_i))}t}\label{e:term2_bound}
\end{align}
where the last inequality comes from the Chernoff bound. In the derivation of the last inequality in (\ref{e:term2_bound}), we should have $f(\theta_i)<\frac{f(\theta_k)d_1}{d_2+f(\theta_k)}$ for $i>k$. This is satisfied from the following fact
\begin{align*}
&f(\theta_i)(d_2+f(\theta_k))\left(\frac{1-P_f^i}{P_d^i-P_f^i}-\frac{1-P_f^k}{P_d^k-P_f^k}\right)
=f(\theta_i)\left(f(\theta_k)\frac{1-P_f^i}{P_d^i-P_f^i}-\theta_k(1-P_f^k)\right)
\\&{<}
f(\theta_k)f(\theta_i)\left(\frac{1-P_f^i}{P_d^i-P_f^i}-\frac{\theta_i(1-P_f^i)}{f(\theta_i)}\right)=f(\theta_k)\left(\frac{1-P_f^i}{P_d^i-P_f^i}-\frac{1-P_f^k}{P_d^k-P_f^k}\right)d_1
\end{align*}
where the first equality comes from the definition of $d_2$, the inequality comes from $\frac{(1-P_f^i)\theta_i}{f(\theta_i)}<\frac{(1-P_f^k)\theta_k}{f(\theta_k)}$ for $i>k$, and the last equality comes from the definition of $d_1$.

The second sub-term in (\ref{e:second_term_decom}) is bounded as
{\small \begin{align}
&\int_{-\infty}^{\frac{tf(\theta_k)d_1}{d_2+f(\theta_k)}} g_{X_i^t}(x)\int_{-\infty}^{\frac{td_2x}{td_1-x}} g_{X_k^t}(y)\,\mathrm{d}y\mathrm{d}x\nonumber
\\&
\overset{\substack{R_i=f(\theta_i)(1-f(\theta_i))t\\R_k=f(\theta_k)(1-f(\theta_k))t}}{\le}\int_{-\infty}^{\frac{tf(\theta_k)d_1}{d_2+f(\theta_k)}}\frac{1}{2\sqrt{2\pi} \sqrt {R_i}}e^{-\frac{(x-f(\theta_i) t)^2}{2R_i}}e^{-\frac{\left(\frac{td_2x}{td_1-x}-f(\theta_k)t\right)^2}{2R_k}}\mathrm{d}x\nonumber
\\&
\overset{(a)}{\le}\int_{-\infty}^{\frac{tf(\theta_k)d_1}{d_2+f(\theta_k)}}\frac{1}{2\sqrt {2\pi} \sqrt {R_i}}e^{-\frac{\left(x-f(\theta_i) t\right)^2}{2R_i}}e^{-\frac{\left(\frac{d_2+f(\theta_k)}{d_1}x-f(\theta_k)t\right)^2}{2R_k}}\mathrm{d}x\nonumber
\\&
\overset{\substack{A=(d_2+f(\theta_k))^2/d_1^2\\H=tf(\theta_k)d_1/(d_2+f(\theta_k))}}{=}\int_{-\infty}^{\frac{tf(\theta_k)d_1}{d_2+f(\theta_k)}}\frac{1}{2\sqrt {2\pi} \sqrt {R_i}}e^{-\frac{R_k(x-tf(\theta_i))^2+R_iA(x-H)^2}{2R_iR_k}}\mathrm{d}x\nonumber
\\&
=e^{-\frac{(R_k+R_iA)\left(R_kt^2f(\theta_i)^2+R_iAH^2\right)-\left(R_ktf(\theta_i)+R_iAH\right)^2}{(R_k+R_iA)2R_iR_k}}\int_{-\infty}^{\frac{tf(\theta_k)d_1}{d_2+f(\theta_k)}}\frac{1}{2\sqrt {2\pi} \sqrt {R_i}}e^{-\frac{(R_k+R_iA)\left(x-\frac{R_ktf(\theta_i)+R_iAH}{R_k+R_iA}\right)^2}{2R_iR_k}}\mathrm{d}x\nonumber
\\&
=
\frac{1}{2}\sqrt{\frac{R_k}{R_k+R_iA}}e^{-\frac{(R_k+R_iA)\left(R_kt^2f(\theta_i)^2+R_iAH^2\right)-\left(R_ktf(\theta_i)+R_iAH\right)^2}{(R_k+R_iA)2R_iR_k}}
Q\left(\frac{\frac{R_ktf(\theta_i)+R_iAH}{R_k+R_iA}-tf(\theta_i)}{\sqrt{\frac{R_iR_k}{R_k+R_iA}}}\right)\nonumber
\\&
\le
\frac{1}{4}\sqrt{\frac{R_k}{R_k+R_iA}}e^{-\frac{(R_k+R_iA)\left(R_kt^2f(\theta_i)^2+R_iAH^2\right)-\left(R_ktf(\theta_i)+R_iAH\right)^2+\left(tf(\theta_i)-\frac{R_ktf(\theta_i)+R_iAH}{R_k+R_iA}\right)^2(R_k+R_iA)^2}
{(R_k+R_iA)2R_iR_k}}\nonumber
\\&
\overset{(b)}{\le}
\frac{1}{4}\sqrt{\frac{R_k}{R_k+R_iA}}e^{-\frac{t^2(f(\theta_i))^2(R_k+R_iA)}{2R_iR_k}}=
\frac{1}{4}\sqrt{\frac{R_k}{R_k+R_iA}}e^{-\frac{(f(\theta_i))^2\left(f(\theta_k)(1-f(\theta_k))+f(\theta_i)(1-f(\theta_i))A\right)}
{2f(\theta_i)(1-f(\theta_i))f(\theta_k)(1-f(\theta_k))}t}\label{e:term3_bound}
\end{align}}
where $(a)$ comes from the fact that for $x\in(-\infty,\frac{tf(\theta_k)d_1}{d_2+f(\theta_k)}]$, we have $\frac{td_2x}{td_1-x}\le\frac{d_2+f(\theta_k)}{d_1}x\le tf(\theta_k)$, $(b)$ comes from the fact that since $f(\theta_i)<d_1$ we have $tf(\theta_i)<H \overset{\triangle}{=}tf(\theta_k)d_1/(d_2+f(\theta_k))$, and other inequalities come from the Chernoff bound.

From (\ref{e:algo2_false_prob}) and (\ref{e:single_het_prob_rewrite})-(\ref{e:term3_bound}), we can bound the false access probability, for Slot $t$ when $t\ge t_0$, as
\begin{align*}
&\text{Prob}\left( k_{\hat{\mathbf{\Theta}}(t)} \left(\textbf{X}(t)\right)\ne k_{\mathbf{\Theta}}\left(\textbf{X}(t)\right)\right)\le
\sum\limits_{u\in {\cal U}}\sum\limits_{i>k, i\in {\cal I}_u, k\in {\cal I}_u} \text{Prob}\left(k_{\hat{\mathbf{\Theta}}(t)}(u)\ne k_{\mathbf{\Theta}}(u)\right)\text{Prob}\left(\textbf{X}(t)=u\right) \\&
\le \sum\limits_{u\in {\cal U}} \sum\limits_{i>k, i\in {\cal I}_u, k\in {\cal I}_u} \Bigg(\frac{1}{2}e^{-\frac{(d_1-f(\theta_i))^2}
{2f(\theta_i)(1-f(\theta_i))}t}+\frac{1}{2}e^{-\frac{\left(\frac{f(\theta_k)d_1}{d_2+f(\theta_k)}-f(\theta_i)\right)^2}{2f(\theta_i)(1-f(\theta_i))}t}
\\&
+\frac{1}{4}\sqrt{\frac{R_k}{R_k+R_iA}}e^{-\frac{f(\theta_i)^2\left(f(\theta_k)(1-f(\theta_k))+f(\theta_i)(1-f(\theta_i))A\right)}
{2f(\theta_i)(1-f(\theta_i))f(\theta_k)(1-f(\theta_k))}t}\Bigg)\text{Prob}\left(\textbf{X}(t)=u\right)
\\&
\le c_3e^{-c_4t}
\end{align*}
where $c_3\!=\!\frac{5}{4}{|{\cal U}| \choose 2}$, $c_4\!=\!\min\limits_{i>k}\left\{\frac{(d_1-f(\theta_i))^2}
{2f(\theta_i)(1-f(\theta_i))},\frac{\left(\frac{f(\theta_k)d_1}{d_2+f(\theta_k)}-f(\theta_i)\right)^2}{2f(\theta_i)(1-f(\theta_i))},\frac{f(\theta_i)^2\left(f(\theta_k)(1-f(\theta_k))+f(\theta_i)(1-f(\theta_i))A\right)}
{2f(\theta_i)(1-f(\theta_i))f(\theta_k)(1-f(\theta_k))}\right\}>0$.

Therefore, for regret $R(t)$ of Algorithm \ref{algo2}, we have
\begin{align}
&\limsup\limits_{t\to\infty}R(t)
\le \limsup\limits_{t\to\infty}\sum\limits_{j=1}^t c_0\text{Prob}\left( k_{\hat{\mathbf{\Theta}}(j)} \left(\textbf{X}(j)\right)\ne k_{\mathbf{\Theta}}\left(\textbf{X}(j)\right)\right)\nonumber
\\&
\le \sum\limits_{j=1}^{t_0} c_0\text{Prob}\left( k_{\hat{\mathbf{\Theta}}(j)} \left(\textbf{X}(j)\right)\ne k_{\mathbf{\Theta}}\left(\textbf{X}(j)\right)\right)+
\limsup\limits_{t\to\infty}\sum\limits_{j=t_0+1}^t c_0\text{Prob}\left( k_{\hat{\mathbf{\Theta}}(j)} \left(\textbf{X}(j)\right)\ne k_{\mathbf{\Theta}}\left(\textbf{X}(j)\right)\right)\nonumber
\\&
\le c_0t_0+\sum\limits_{j=t_0+1}^\infty c_0c_3e^{-c_4j}<\infty\label{e:finite_algo2}.
\end{align}

\section{Proof of Theorem \ref{th:case2_single_order_opt}}\label{a:case2_single_order_opt}
Recall that we assume $\theta_1>\theta_2>...>\theta_N$, and for the genie-aided rule, $\mathcal{M}^*=\{1,2,...,M\}$ is the optimal set of channels to sense. Then for any rule, the expected reward loss in a slot (say Slot $j$) is bounded by the  maximal expected reward of the genie-aided rule in the slot, given as $\Delta\overset{\triangle}{=}B(T-\tau)E\big[\max\limits_{i\in {\cal M}^*}\frac{\theta_i(1-P_f)}{f(\theta_i)}X_i(j)\big]$, where ${f(\theta_i)}=(1-P_{f}^i)\theta_i+(1-P_d^i)(1-\theta_i)$ is the probability that Channel $i$ is sensed free. Throughout our proofs, $I_{\{{\cal{A}}\}}$ is an indicator function for an event ${\cal A}$.

Recall that in Algorithm \ref{algo3}, ${\cal M}(j)$ denotes the set of channels to sense at Slot $j$. So until Slot $t$, the regret $R(t)$ of Algorithm \ref{algo3} is bounded as
\begin{align}
&R(t)
\le\Delta\sum\limits_{j=1}^t
E\left[I_{\left\{\mathcal{M}(j)\ne \mathcal{M}^*\right\}}\right] \nonumber \\&+\Delta\sum\limits_{j=1}^t
E\Big[I_{\left\{\mathcal{M}(j)=\mathcal{M}^*\right\}}I_{\left\{\mathop\cup \limits_{i<k,i\in
{\cal I}_{{\cal M}^*}(j),k\in
{\cal I}_{{\cal M}^*}(j)}
\left\{\hat \theta_i(j)+\frac{1}{P_d-P_f}\sqrt {\frac{2\ln (j-1)}{T_i(j-1)}}<\hat\theta_k(j)+\frac{1}{P_d-P_f}\sqrt {\frac{2\ln (j-1)}{T_k(j-1)}}\right\}
\right\}}\Big]\label{e:bound_r_t}
\end{align}
where ${\cal I}_{{\cal M}^*}(j)$ denotes sensed-free channels in Slot $j$ when channels in ${\cal M}^*$ are sensed. On the right hand side of (\ref{e:bound_r_t}), the first term is the regret bound when the secondary user does not select exactly ${\cal M}^*$ to sense (i.e., ${\cal M}(j)\neq {\cal M}^*$), and the second term is the regret bound when the secondary user senses channels in ${\cal M}^*$ but does not select the best sensed-free channel to access.

In the sequel of this proof, for Slot $j$, denote $\hat \theta^T_k(T_k(j-1))$ as the estimated free probability of Channel $k$, as described in Algorithm \ref{algo3}, when Channel $k$ has been sensed by $T_k(j-1)$ slots until Slot $j-1$.

Now we derive a bound for the first term on the right hand side of (\ref{e:bound_r_t}). Recall that $T_i(t)$ is the number of slots in which Channel $i$ is sensed until Slot $t$. Then we have
\begin{equation}\label{e:first_term_theorem4_bound}
\sum\limits_{j=1}^t
E\left[I_{\left\{\mathcal{M}(j)\ne \mathcal{M}^*\right\}}\right] \le \sum\limits_{i=M+1}^N E[T_i(t)].
\end{equation}
Further, for $M+1\le i \le N$ and any positive integer $l$, we have
\begin{equation}
\begin{array}{lll}
T_i(t)&=1+\sum\limits_{j=\left\lceil \frac{N}{M} \right\rceil+1}^t I_{\left\{i\in \mathcal{M}(j)\right\}}=
1+\sum\limits_{j=\left\lceil \frac{N}{M} \right\rceil+1}^t I_{\left\{i\in \mathcal{M}(j),~T_i(j-1)\ge l\right\}}+\sum\limits_{j=\left\lceil \frac{N}{M} \right\rceil+1}^tI_{\left\{i\in \mathcal{M}(j),~T_i(j-1)< l\right\}}
\\&\le l+\sum\limits_{j=\left\lceil \frac{N}{M} \right\rceil+1}^t I_{\left\{i\in \mathcal{M}(j),T_i(j-1)\ge l\right\}}
\\
&\le l+\sum\limits_{j=\left\lceil \frac{N}{M} \right\rceil+1}^t I_{\left\{\min\limits_{k\in{\cal M}^*} \big\{\hat \theta^T_k(T_k(j-1))+\frac{1}{P_d-P_f}\sqrt {\frac{2\ln (j-1)}{T_k(j-1)}}\big\}\le\hat \theta^T_i(T_i(j-1))+\frac{1}{P_d-P_f}\sqrt {\frac{2\ln (j-1)}{T_i(j-1)}}
,~T_i(j-1)\ge l\right\}}
\\&\le
l+\sum\limits_{k=1}^M \sum\limits_{j=\left\lceil \frac{N}{M} \right\rceil}^{t-1} I_{\left\{\hat \theta^T_k(T_k(j))+\frac{1}{P_d-P_f}\sqrt {\frac{2\ln{j}}{T_k(j)}}\le\hat \theta^T _i(T_i(j))+\frac{1}{P_d-P_f}\sqrt {\frac{2\ln j}{T_i(j)}},~T_i(j)\ge l\right\}}
\\&\le
l+\sum\limits_{k=1}^M \sum\limits_{j=\left\lceil \frac{N}{M} \right\rceil}^{t-1}  I_{\left\{\min\limits_{0<t_1\le j} \big\{\hat\theta^T_k(t_1)+\frac{1}{P_d-P_f}\sqrt {\frac{2\ln j}{t_1}}\big\}\le
\max\limits_{l\le t_2\le j} \big\{\hat\theta^T_i(t_2)+\frac{1}{P_d-P_f}\sqrt {\frac{2\ln j}{t_2}}\big\}\right\}}
\\&\le
l+\sum\limits_{k=1}^M \sum\limits_{j=1}^t \sum\limits_{t_1=1}^{j} \sum\limits_{t_2=l}^{j} I_{\left\{\hat\theta^T_k(t_1)+\frac{1}{P_d-P_f}\sqrt {\frac{2\ln j}{t_1}}\le \hat\theta^T_i(t_2)+\frac{1}{P_d-P_f}\sqrt {\frac{2\ln j}{t_2}}\right\}}.\label{e:inferior_time_bound}
\end{array}
\end{equation}

Similar to analysis in \cite{PAuer}, we have the fact that if event $\hat\theta^T_k(t_1)+\frac{1}{P_d-P_f}\sqrt {\frac{2\ln j}{t_1}}\le \hat\theta^T_i(t_2)+\frac{1}{P_d-P_f}\sqrt {\frac{2\ln j}{t_2}}$ happens, then at least one of the following three events will happen: $\hat\theta^T_k(t_1)\le\theta_k-\frac{1}{P_d-P_f}\sqrt {\frac{2\ln j}{t_1}}$, $\hat\theta^T_i(t_2)\ge\theta_i+\frac{1}{P_d-P_f}\sqrt {\frac{2\ln j}{t_2}}$, and $\theta_k<\theta_i+$ $\frac{2}{P_d-P_f}\sqrt {\frac{2\ln j}{t_2}}$. In other words, we have
\begin{equation}\label{e:3prob}
\begin{array}{lll}
&E\left[I_{\Big\{\hat\theta^T_k(t_1)+\frac{1}{P_d-P_f}\sqrt {\frac{2\ln j}{t_1}}\le \hat\theta^T_i(t_2)+\frac{1}{P_d-P_f}\sqrt {\frac{2\ln j}{t_2}}\Big\}}\right]
\\&
\le E\left[I_{\Big\{\hat\theta^T_k(t_1)\le\theta_k-\frac{1}{P_d-P_f}\sqrt {\frac{2\ln j}{t_1}}\Big\}}\right]+E\left[I_{\Big\{\hat\theta^T_i(t_2)\ge\theta_i+\frac{1}{P_d-P_f}\sqrt {\frac{2\ln j}{t_2}}\Big\}}\right]+E\left[I_{\Big\{\theta_k<\theta_i+\frac{2}{P_d-P_f}\sqrt {\frac{2\ln j}{t_2}}\Big\}}\right].
\end{array}
\end{equation}
Using Chernoff-Hoeffding bound, the first two terms on the right hand side of (\ref{e:3prob}) are bounded as
\begin{equation}\label{e:prob1}
E\left[I_{\left\{\hat\theta^T_k(t_1)\le\theta_k-\frac{1}{P_d-P_f}\sqrt {\frac{2\ln j}{t_1}}\right\}}\right]\le j^{-4},
~E\left[I_{\left\{\hat\theta^T_i(t_2)\ge\theta_i+\frac{1}{P_d-P_f}\sqrt {\frac{2\ln j}{t_2}}\right\}}\right]\le j^{-4}.
\end{equation}

We note that if $t_2\ge \frac{8\ln t}{(\theta_M-\theta_i)^2 (P_d-P_f)^2}$, then we always have $\theta_k\ge\theta_i+\frac{2}{P_d-P_f}\sqrt {\frac{2\ln j}{t_2}}$ for any $k \in {\cal M}^*$ and $j\le t$, which means $I_{\Big\{\theta_k<\theta_i+\frac{2}{P_d-P_f}\sqrt {\frac{2\ln j}{t_2}}\Big\}}=0$. Therefore, by setting $l=\left\lceil\frac{8\ln t}{(\theta_M-\theta_i)^2(P_d-P_f)^2}\right\rceil$, from (\ref{e:first_term_theorem4_bound})-(\ref{e:prob1}) we have
\begin{equation}\label{e:bound_fir_term}
\begin{array}{lll}
\sum\limits_{j=1}^t
E\left[I_{\left\{\mathcal{M}(j)\ne \mathcal{M}^*\right\}}\right]   &\le& \sum\limits_{i=M+1}^N \left\lceil {\frac{8\ln t}{(\theta_M-\theta_i)^2(P_d-P_f)^2}}\right\rceil +\sum\limits_{i=M+1}^N \sum\limits_{k=1}^M \sum\limits_{j=1}^\infty \sum\limits_{t_1=1}^{j} \sum\limits_{t_2=\left\lceil {\frac{8\ln t}{(\theta_M-\theta_i)^2(P_d-P_f)^2}}\right\rceil}^j 2j^{-4} \\
&\le & \sum\limits_{i=M+1}^N  {\frac{8\ln t}{(\theta_M-\theta_i)^2(P_d-P_f)^2}} + (N-M)\left(\frac{M\pi^2}{3}+1\right).
\end{array}
\end{equation}

To bound the second term on the right hand side of (\ref{e:bound_r_t}), we have
{\small \begin{equation}\label{e:bound_term2}
\begin{array}{lll}
&\sum\limits_{j=1}^t
I_{\left\{\mathcal{M}(j)=\mathcal{M}^*\right\}}I_{\left\{\mathop\cup \limits_{i<k,i\in
{{\cal I}_{{\cal M}^*}}(j),k\in
{{\cal I}_{{\cal M}^*}}(j)}
\left[\hat \theta^T_i(T_i(j-1))+\frac{1}{P_d-P_f}\sqrt {\frac{2\ln (j-1)}{T_i(j-1)}}<\hat\theta^T_k(T_k(j-1))+\frac{1}{P_d-P_f}\sqrt {\frac{2\ln (j-1)}{T_k(j-1)}}\right]
\right\}}
\\&\le
1\!+\!\!\sum\limits_{j=\left\lceil \frac{N}{M} \right\rceil+1}^t\!\!I_{\left\{\mathcal{M}(j)=\mathcal{M}^*\right\}}I_{\left\{\mathop\cup \limits_{i<k,i,k\in
{{\cal I}_{{\cal M}^*}}(j)}
\left[\hat \theta^T_i(T_i(j-1))+\frac{1}{P_d-P_f}\sqrt {\frac{2\ln (j-1)}{T_i(j-1)}}<\hat\theta^T_k(T_k(j-1))+\frac{1}{P_d-P_f}\sqrt {\frac{2\ln (j-1)}{T_k(j-1)}}\right]
\right\}}
\\&\le
1+\sum\limits_{i<k,~ i,k\in {\cal M}^*}\sum\limits_{j=\left\lceil \frac{N}{M} \right\rceil+1}^tI_{\left\{\mathcal{M}(j)=\mathcal{M}^*\right\}}I_{
\left\{\hat \theta^T_i(T_i(j-1))+\frac{1}{P_d-P_f}\sqrt {\frac{2\ln (j-1)}{T_i(j-1)}}<\hat\theta^T_k(T_k(j-1))+\frac{1}{P_d-P_f}\sqrt {\frac{2\ln (j-1)}{T_k(j-1)}}\right\}}
\\&\le
\sum\limits_{i<k,~i,k\in {\cal M}^*}\Bigg(l_{i,k}+\sum\limits_{j=\left\lceil \frac{N}{M} \right\rceil+1}^t \bigg(I_{\left\{\mathcal{M}(j)=\mathcal{M}^*,T_k(j-1)\ge l_{i,k}\right\}} \\
&~~~~~~~~~~~~~~~~~~~~~~~~~~~~~\cdot I_{
\left\{\hat \theta^T_i(T_i(j-1))+\frac{1}{P_d-P_f}\sqrt {\frac{2\ln (j-1)}{T_i(j-1)}}<\hat\theta^T_k(T_k(j-1))+\frac{1}{P_d-P_f}\sqrt {\frac{2\ln (j-1)}{T_k(j-1)}}\right\}}\bigg)\Bigg)
\\&\le
\sum\limits_{i<k,~i,k\in {\cal M}^*}\Bigg\{l_{i,k}+\sum\limits_{j=1}^t\sum \limits_{t_1=1}^j \sum\limits_{t_2=l_{i,k}}^j I_{\left\{\hat \theta^T _i(t_1)+\frac{1}{P_d-P_f}\sqrt {\frac{2\ln j}{t_1}}<\hat\theta^T_k(t_2)+\frac{1}{P_d-P_f}\sqrt {\frac{2\ln j}{t_2}}\right\}}\Bigg\}
\end{array}
\end{equation}}
where $l_{i,k}$ can be an arbitrary positive integer.

Similar to the treatments in (\ref{e:3prob})-(\ref{e:bound_fir_term}), the second term on the right hand side of (\ref{e:bound_r_t}) is bounded as
{\it \begin{equation}\label{e:part_2_b}
\begin{array}{lll}
&\Delta\sum\limits_{j=1}^t E\left[I_{\left\{\mathcal{M}(j)=\mathcal{M}^*\right\}}I_{\left\{\mathop\cup \limits_{i<k,i,k\in
{{\cal I}_{{\cal M}^*}}(j)}
\left\{\hat \theta^T_i(T_i(j-1))+\frac{1}{P_d-P_f}\sqrt {\frac{2\ln (j-1)}{T_i(j-1)}}<\hat\theta^T_k(T_k(j-1))+\frac{1}{P_d-P_f}\sqrt {\frac{2\ln (j-1)}{T_k(j-1)}}\right\}
\right\}}\right]\\&
\\&
\le \Delta \ln t\sum\limits_{i<k\in {\cal M}^*}\frac{8}{(\theta_i-\theta_k)^2(P_d-P_f)^2}+\Delta {M \choose 2}\big(\frac{\pi^2}{3}+1\big).
\end{array}
\end{equation}}
Then, from (\ref{e:bound_r_t}), (\ref{e:bound_fir_term}) and (\ref{e:part_2_b}), the regret until Slot $t$, $R(t)$, is bounded as
{\small \begin{multline}
R(t)
\le\Delta
\ln t\sum\limits_{i=M+1}^N\frac{8}{(\theta_M-\theta_i)^2(P_d-P_f)^2}
+\Delta\ln t\sum\limits_{i<k\in {\cal M}^*}\frac{8}{(\theta_i-\theta_k)^2(P_d-P_f)^2}\\+\Delta (N-M)\left(\frac{M\pi^2}{3}+1\right)+\Delta {M \choose 2}\left(\frac{\pi^2}{3}+1\right).\label{e:bound_all}
\end{multline}}
In other words, $R(t)\sim O(\ln t)$ for finite $t$ and for $t\rightarrow \infty$.

\section{Proof of Theorem \ref{th:case2_general_order_opt}}\label{a:case2_general_order_opt}

Denote ${\cal M}_{i^*}$ as the optimal set of channels to sense (i.e., the set of channels to sense in the genie-aided rule). Denote ${\cal M}(t)$ as the channel set decided by Algorithm \ref{algo4} to be sensed at Slot $t$. Similar to proof of Theorem \ref{th:case2_single_order_opt}, the regret $R(t)$ until Slot $t$ is bounded as
\begin{multline}\label{e:bound_all2}
R(t)
\le\Delta\sum\limits_{j=1}^t
E\left[I_{\left\{\mathcal{M}(j)\ne \mathcal{M}_{i^*}\right\}}\right]+\Delta\sum\limits_{j=1}^t
E\Bigg[I_{\left\{\mathcal{M}(j)=\mathcal{M}_{i^*}\right\}}\\\cdot I_{\left\{\mathop\cup \limits_{ \overset{m_{i^*,k}, m_{i^*,r}\in
{\cal I}_{{\cal M}_{i^*}}(j)}{E[S_{m_{i^*,k}}|X_{m_{i^*,k}}=1] > E[S_{m_{i^*,r}}|X_{m_{i^*,r}}=1]    }   }
\left\{
\frac{Y_{i^*,k}(j-1)}{T_{i^*,k}(j-1)}+\sqrt{ \frac{2\ln (j-1)}{T_{i^*,k}(j-1)}}<
\frac{Y_{i^*,r}(j-1)}{T_{i^*,r}(j-1)}+\sqrt{ \frac{2\ln (j-1)}{T_{i^*,r}(j-1)}}
\right\}
\right\}}\Bigg].
\end{multline}
Next we derived bounds for the two terms on the right hand side of (\ref{e:bound_all2}), respectively.

Since $T_i(t)$ is the number of slots that channel set ${\cal M}_i$ is sensed until Slot $t$, the first term on the right hand side of (\ref{e:bound_all2}) is
$\Delta \sum\limits_{j=1}^t
E\left[I_{\left\{\mathcal{M}(j)\ne \mathcal{M}^*\right\}}\right]=\Delta\sum\limits_{{i\ne i^*,i\in\{1,2,...,{N\choose M}\}}}E[T_i(t)].$

For each $i\in\{1,2,...,{N\choose M}\}$, it can be proved that the reward sequence $Y_i(t)|_{T_i(t)=1}$, $Y_i(t)|_{T_i(t)=2}$, ..., $Y_i(t)|_{T_i(t)=n}$ satisfy a so-called {\it drift condition}\footnote{Its definition is given in Section 2.4 of \cite{Kocsis}.}. The proof is omitted due to space limit. 

Similar to the treatments in (\ref{e:inferior_time_bound})-(\ref{e:bound_fir_term}), 
we have
$E[T_i(t)]\le \frac{8\ln t}{\xi_i}+\frac{\pi^2}{3}+1$
where \[\xi_i\overset{\triangle}{=}\Big(E\Big[\max\limits_{l\in{\cal I}_{{\cal M}_{i^*}} } E\left[S_l|X_l=1\right]\Big]-E\Big[\max\limits_{l\in{\cal I}_{{\cal M}_i} } E\left[S_l|X_l=1\right]\Big]\Big)^2\] and ${\cal I}_{{\cal M}_i}$ is the set of sensed-free channels if ${\cal M}_i$ is sensed.
Therefore, the first term on the right hand side of (\ref{e:bound_all2}) is bounded as
\begin{equation}\label{e:term1_b2}
\Delta\sum\limits_{j=1}^t
E\left[I_{\left\{\mathcal{M}(j)\ne \mathcal{M}^*\right\}}\right]\le \Delta \ln t \sum\limits_{\substack{i\in\{1,2,...,{N \choose M}\}\\i\ne i^*}}\frac{8}{\xi_i}  +\Delta \left({N \choose M}-1\right)\left(\frac{\pi^2}{3}+1\right).
\end{equation}

Similar to the treatments in (\ref{e:bound_term2})-(\ref{e:part_2_b}), we have a bound for the second term on the right hand side of (\ref{e:bound_all2}) as $\Delta \ln t\sum\limits_{k<r\le M}\frac{8}{\left(\frac{\left(1-P_f^{m_{i^*,k}}\right)\theta_{m_{i^*,k}}}{f(\theta_{m_{i^*,k}})}-\frac{\left(1-P_f^{m_{i^*,r}}\right)\theta_{m_{i^*,r}}}{f(\theta_{m_{i^*,r}})}\right)^2}+\Delta {M \choose 2}\big(\frac{\pi^2}{3}+1\big)$.

It can be seen that, the two terms on the right hand side of (\ref{e:bound_all2}) are bounded by $O(\ln t)$. Therefore, the regret until Slot $t$, $R(t)$, is $O(\ln t)$.


\end{document}